\documentclass[%
 reprint,
nofootinbib,
 amsmath,amssymb,
 aps,
 pre,
]{revtex4-1}

\usepackage{graphicx}
\usepackage{dcolumn}
\usepackage{bm}

\usepackage[utf8]{inputenc}
\usepackage{xcolor}
\usepackage[normalem]{ulem}

\usepackage[colorlinks=true, allcolors=blue]{hyperref}
\usepackage{multirow}

\usepackage{scalerel,stackengine}
\stackMath
\newcommand\reallywidehat[1]{%
\savestack{\tmpbox}{\stretchto{%
  \scaleto{%
    \scalerel*[\widthof{\ensuremath{#1}}]{\kern-.6pt\bigwedge\kern-.6pt}%
    {\rule[-\textheight/2]{1ex}{\textheight}}
  }{\textheight}%
}{0.5ex}}%
\stackon[1pt]{#1}{\tmpbox}%
}

\newcommand{\V}[1]{\mathbf{#1}}
\newcommand{\shellK}{\mathrm{K}}
\newcommand{\shellQ}{\mathrm{Q}}
\newcommand{\Vshell}[2]{\mathbf{#1}^{\mathrm{#2}}}
\newcommand{\shell}[2]{{#1}^{\mathrm{#2}}}
\newcommand{\FT}[1]{\reallywidehat{#1}}

\newcommand{\T}[1]{\mathcal{T}_{\mathrm{#1}}}

\newcommand{\RE}[1]{\mathrm{Re} \left [ #1 \right ]}
\newcommand{\bra}[1]{\left ( #1 \right )}

\newcommand{\pd}{\partial}
\newcommand{\Eint}{E_\mathrm{int}}
\newcommand{\Ekin}{E_\mathrm{kin}}
\newcommand{\Etot}{E_\mathrm{tot}}

\newcommand{\imagi}{\mathrm{i}}
\newcommand{\Eulere}{\mathrm{e}}

\newcommand{\cs}{c_{\mathrm{s}}}

\newcommand{\dd}{\mathrm{d}}

\usepackage{cancel}

\def\apj{ApJ}

\def\aap{A\&A}

\def\mnras{MNRAS}

\def\pre{Phys.~Rev.~E}
\def\prl{Phys.~Rev.~Lett.}

\def\physrep{Phys.~Rep.}

\begin{document}

\preprint{APS/123-QED}

\title[Energy transfer in forced compressible turbulence]{
	Kinetic and internal energy transfer in implicit large eddy simulations of
	forced compressible turbulence
}

\author{Wolfram Schmidt}
\email{wolfram.schmidt@uni-hamburg.de}
\affiliation{
Hamburger Sternwarte, Universität Hamburg, Gojenbergsweg 112, D-21029 Hamburg, Germany
}

\author{Philipp Grete}
\email{grete@pa.msu.edu.}
\affiliation{
Department of Physics and Astronomy,
Michigan State University, East Lansing, MI 48824, USA}

\date{\today}

\begin{abstract}
We revisit the problem of how energy transfer through the turbulent cascade operates in compressible hydrodynamic turbulence.
In general, there is no conservative compressible cascade since the kinetic and internal energy reservoirs can exchange energy through pressure dilatation. Moreover, statistically stationary turbulence at high Mach number can only be maintained in nearly isothermal gas, i.e.\ if excess heat produced by shock compression and kinetic energy dissipation is continuously removed from the system. We mimic this process by a linear cooling term in numerical simulations of turbulence driven by stochastic forcing. This allows us to investigate turbulence statistics for a broad range of Mach numbers. We compute the rate of change of kinetic and internal energy in wavenumber shells caused by advective, compressive, and pressure dilatation effects and constrain power-law fits to compressible turbulence energy spectra to a range of wavenumbers in which the total energy transfer is close to zero. The resulting scaling exponents are significantly affected by the forcing. Depending on the root mean square Mach number, we find a nearly constant advective component of the cross-scale flux of kinetic energy at intermediate wavenumbers for particular mixtures of solenoidal and compressive modes in the forcing. This suggests the existence of a natural, Mach number dependent mixture of forcing modes. Our findings also support an advection-dominated regime at high Mach numbers with specific scaling exponents (Burgers scaling for the pure velocity fluctuation $u$ and Kolmogorov scaling for the mass-weighted variable $v=\rho^{1/3}u$).
\end{abstract}

\pacs{Valid PACS appear here}
\maketitle

\section{Introduction}

Kolmogorov's theory of incompressible isotropic turbulence states that the flux of kinetic energy across a given wavenumber is constant for wavenumbes in the inertial subrange in which both large-scale features of the flow (stirring, boundary conditions, etc.) and viscous dissipation are negligible \cite{Frisch1995,Alexakis2018}. An important consequence is the two-thirds law for the scaling of the squared velocity fluctuation, corresponding to the famous $k^{-5/3}$ energy spectrum. However, it has proven to be difficult to carry over the concepts of this theory to compressible turbulence. A key problem is that, in general, kinetic energy does not cascade conservatively because it can be converted into internal energy and vice versa through pressure dilatation. An important step in addressing this problem was the demonstration of scale-locality for the transfer of kinetic energy in compressible turbulence by Aluie \cite{Aluie2011}. Under certain assumptions (sufficiently fast decline of the pressure dilatation cospectrum), it can be shown that the exchange of kinetic and internal energy becomes subdominant and the two energy budgets are decoupled on sufficiently small scales. In this case, a compressible turbulent cascade with constant kinetic energy flux can exist. At about the same time, the generalization of the relation between the energy flux and two-point-correlation functions for the fully compressible case was analytically derived by Galtier \& Banerjee \cite{Galtier2011}. In addition to a term that reduces to the Kolmogorov result in the incompressible limit, there are contributions related to compression effects (non-zero divergence of the flow) and deviations from a conservative kinetic energy cascade.

A further difficulty is the ambiguity in the scale-decomposition of energy density and fluxes in the compressible case. For example, the energy density in spectral space can be based on the Fourier transforms of the primitive velocity variable $u$ and momentum density $\rho u$ \cite{Graham2010}. Alternatively, the energy density can be considered as square of the variable $w=\sqrt{\rho}\,u$ \cite{Kida1990} in analogy to the incompressible case (where the energy is simply given by $u^2$). For energy flux, there are even more variants \cite{Alexakis2018}. For example, two-point correlation functions of velocity and momentum increments inspired by the derivation of the four-fifth law are used in \cite{Galtier2011,Wagner2012,Kritsuk2013}, while spatial coarse-graining results in expressions linked to subgrid-scale terms in large eddy simulations \cite{Eyink2009,Aluie2013,Zhao2018}. In addition, different flavors of spectral decomposition are applied to compute energy transfers between scales (basically, the transfer is the rate of change of the energy flux within a range of wave numbers). This is the most common method to analyze numerical data \cite{Alexakis2005,Graham2010,Mininni2011,Grete2017a,Verma2019}.

Numerical experiments play a crucial role in studying the phenomenology of compressible turbulence. While it remains challenging to produce strongly compressible or even supersonic turbulent flows under laboratory conditions, the regime of high Mach numbers is accessible in massively parallelized simulations on supercomputers. In astrophysics, substantial interest into supersonic turbulence was triggered by the theoretical investigation of star formation in interstellar gas clouds \cite{MacLow2004,Hennebelle2012}. In a landmark work on scaling properties of supersonic isothermal turbulence \cite{Kritsuk2007} the hypothesis was put forward that mass-weighing of the velocity with $\rho^{1/3}$ allows for the extension of Kolmogorov scaling into the compressible regime. This idea was to some extent questioned in \cite{Schmidt2008,Federrath2010,Federrath2013}. However, the results of these studies remained inconclusive, as only particular Mach numbers were investigated and turbulence was driven by either solenoidal or compressive (dilatational) forcing modes. Further examples for numerical studies of driven hydrodynamic turbulence are \cite{Schmidt2009,Price2011,Aluie2012,Konstandin2012b,Wagner2012,Kritsuk2013}. 

In this article, we pick up ideas formulated in \cite{Kida1990,Alexakis2005} to compute the shell-to-shell transfer of kinetic energy in compressible turbulence based on the velocity variable $w=\sqrt{\rho}\,u$ (Section~\ref{sec:theory}). By defining $q=\sqrt{\rho}\,\cs$, where $\cs$ is the speed of sound, we incorporate internal energy transfers along the same lines as for magnetic energy transfers in compressible magnetohydrodynamic (MHD) turbulence with $B=\sqrt{\rho}\,v_{\rm a}$ ($v_{\rm a}$ is the Alfv\'enic velocity) as basic variable for the spectral decomposition \cite{Grete2017a}. This also guarantees positivity of the spectral energy densities. A similar approach was recently put forward by \cite{Mittal2019}. Our goal is to investigate whether there is a compressible inertial subrange defined by zero energy transfer (corresponding to constant flux) and how it is affected by pressure-dilatation effects for Mach numbers ranging from transonic to supersonic (Sections~\ref{sec:numericalTransfer} and~\ref{sec:crossFlux}). Moreover, we investigate the impact of different mixtures of solenoidal and compressive modes in the forcing. Based on our results we attempt to determine scaling exponents of turbulent energy spectra with different mass-weighing (Section~\ref{sec:numericalSpect}).

Since time-averaging over many snapshots of the flow is paramount to obtain meaningful statistics, we need to maintain a statistically stationary state over several dynamical timescales. As in most simulations of compressible turbulence mentioned above, we perform implicit large eddy simulations, where dissipation results from numerical truncation terms, instead of explicitly treating viscous dissipation (see \cite{Schmidt2015} and references cited therein).
Explicit physical viscosity (bulk and shear) is neglected since Reynolds 
numbers in typical astrophysical systems are by far too high to allow for
direct numerical simulations. 
Regardless of the nature of dissipation, the mean heating rate due to the dissipation of kinetic energy approximately equals the rate of energy injected by the forcing. To avoid a declining Mach number of supersonic turbulence, excess heat has to be continuously removed from the system. This is commonly achieved in numerical simulations of driven turbulence either by applying an isothermal equation of state (i.e.\ setting the speed of sound $\cs$ indentical to a constant, which reduces the system of dynamic variables to mass density and momentum) or by artificially increasing the internal energy by setting the adiabatic exponent to a value that differs by a tiny fraction from unity (in this case, only the pressure is physical). Both methods are not suitable for our study because we need a physically meaningful (i.e.\ neither constant nor artificially increased) internal energy to compute transfers between the kinetic and internal energy reservoirs. For this reason, we define internal energy on the basis of the ideal gas equation and apply a simple toy model to mimic radiative cooling of the gas with finite cooling time. Details about the numerical simulations are outlined in Section~\ref{sec:stat} and Appendix~\ref{sec:compute}. The outcome of our analysis is discussed in Section~\ref{sec:concl}.

\bigskip

\section{Theory and methods}
\label{sec:theory}

\subsection{Dynamical equations}

The equations of compressible gas dynamics in conservative form are given by
\begin{align}
\label{eq:rho}
\pd_t \rho + \pd_i (\rho u_i) &= 0 \;, \\
\label{eq:rhoU}
\pd_t (\rho u_i) + \pd_j (\rho u_i u_j) + \pd_i p &= f_i \;, \\
\label{eq:rhoe}
\pd_t (\rho e) + \pd_j (\rho u_j e) + p\, \pd_j u_j  &= \Lambda_{\rm HC} \;.
\end{align}
The density is denoted by $\rho$, the velocities by $\V{u}$, the internal energy density by $\rho e$, and the thermal pressure by $p=(\gamma-1)\rho e$, where $\gamma$ is the adiabatic exponent of the gas. The source term $\V{f}=\rho\V{a}$ on the right hand side of the momentum equation~(\ref{eq:rhoU}) represents an external force density and $\Lambda_{\rm HC}$ on the right hand side of the internal energy equation~(\ref{eq:rhoe}) a heating and cooling function (see Section \ref{sec:numericalData} for the modeling of these terms in our simulations).

\subsubsection{Decomposition of kinetic energy}

As discussed in \cite{Grete2017a}, there is no unambiguous definition of the spectral kinetic energy density $\Ekin$ in the compressible case. Since the symmetrized expression $\RE{\FT{u}_i \FT{\rho u}_i^*}/2$ used in \cite{Graham2010} does not guarantee positive definiteness in wavenumber space, we use for our analysis kinetic energy densities defined by
\begin{align}
\int \underbrace{\frac{1}{2} w_i w_i}_{\equiv \Ekin(\V{x})} \mathrm{d}\V{x} = 
\frac{1}{\bra{2\pi}^3} \int \underbrace{\frac{1}{2} \FT{w}_i \FT{w}_i^*}_{\equiv \Ekin(\V{k})} 
\mathrm{d}\V{k}\;,
\end{align}
where $\V{w} \equiv \sqrt{\rho}\, \V{u}$ \cite{Kida1990}.

The corresponding dynamical equations are derived in \citet{Kida1990} for the hydrodynamic case and in \cite{Grete2017a} for the general MHD case. In the hydrodynamic case, the dynamical equation for the kinetic energy density in real space reduces to
\begin{align}
\label{eq:KinEnReal}
\begin{split}
\pd_t \Ekin (\V{x}) = 
&- w_i u_j \pd_j w_i  - \frac{1}{2} w_i w_i \pd_j u_j \\
&- \frac{w_i}{\sqrt{\rho}} \pd_i p + w_i \sqrt{\rho} a_i\;.
\end{split}
\end{align}
An equivalent representation of the energy equation in terms of the mass-weighted variable $\V{v} = \rho^{1/3}\, \V{u}$ cannot be obtained, although this variable seems to be the natural choice for flux terms on the right-hand side \cite{Kritsuk2007}.

\begin{widetext}
\noindent In wavenumber space, the kinetic energy density is given by
\begin{align}
\label{eq:KinEnFT}
\pd_t \Ekin (\V{k}) = \RE{ 
- \FT{w_i} \FT{u_j \pd_j w_i}^*  
- \frac{1}{2} \FT{w_i} \FT{w_i \pd_j u_j}^* 
- \FT{w_i} \FT{\frac{1}{\sqrt{\rho}} \pd_i p}^* 
+ \FT{w_i} \FT{\sqrt{\rho} a_i}^*
}\;,
\end{align}
where wide hats signifiy Fourier transforms and an asterisk complex conjugation.
\end{widetext}

\subsubsection{Decomposition of internal energy}

The internal energy density $\Eint = \rho e$ can be expressed in terms of the speed of sound $\cs$ as
\begin{equation}
\Eint = \frac{p}{\gamma-1} = \frac{\rho\cs^2}{\gamma(\gamma-1)}
\end{equation}
This suggests a decomposition of the internal energy density in terms of the
variable
\begin{equation}
\label{eq:q_def}
q = \sqrt{\rho}\,\cs
\end{equation}
such that
\begin{equation}
\Eint = \frac{q^2}{\gamma(\gamma-1)}
\end{equation}
and total energy can be written as
\begin{equation}
\label{eq:TotEnDef}
\Etot = \Ekin + \Eint = \frac{|\V{w}|^2}{2} + \frac{q^2}{\gamma(\gamma-1)}\;.
\end{equation}

The dynamical equations can be derived in much the same way as for the kinetic energy densities. Starting from equation~(\ref{eq:rhoe}), it is straightforward to show that
\begin{align}
\label{eq:IntEnReal}
\begin{split}
\pd_t q^2 = 
&- 2q u_j \pd_j q  - q^2 \pd_j u_j \\
&- (\gamma-1) q^2 \pd_j u_j + \gamma(\gamma-1) \Lambda_{\rm HC}\;.
\end{split}
\end{align}
This representation differs from the one used in \citep{Kida1990} given that we use $q$
as a building block and allows for a more straightforward interpretation of the terms.
The second term on the right hand side is analogous to the corresponding term in equation~(\ref{eq:KinEnReal}).
Although the second and third term could be combined, we will show in the following that they have different interpretations in the context of energy transfers.

\begin{widetext}
\noindent The representation of the internal energy equation in Fourier space thus reads
\begin{align}
\label{eq:IntEnFT}
\pd_t \Eint (\V{k}) = \RE{ 
- \frac{2}{\gamma(\gamma-1)}\FT{q}\,\FT{u_j \pd_j q}^* - \frac{1}{\gamma(\gamma-1)}\FT{q}\,\FT{q \pd_j u_j}^* 
- \frac{1}{\gamma}\FT{q}\,\FT{q \pd_j u_j}^* + \FT{q}\,\FT{q^{-1}\Lambda_{\rm HC}}^*
}\;.
\end{align}
\end{widetext}

To conclude this section, we consider the isothermal limit $q^2 \rightarrow c_0^2\rho$, where $c_0$ is the constant isothermal sound speed given by $c_0=\sqrt{P/\rho}$ with $\gamma = 1$. Formally, this corresponds to infinite internal energy, as $P/(\gamma-1)$ diverges. However, $q^2$ remains a finite quantity.
In the isothermal limit, equation~\eqref{eq:IntEnReal} becomes
\begin{align}
\label{eq:q_isoth}
\pd_t q^2 = - u_j \pd_j q^2  - q^2 \pd_j u_j \;.
\end{align}
Division by $c_0^2$ yields
\begin{align}
\pd_t \rho = - \pd_j (u_j\rho)\;.
\end{align}
This is, of course, the continuity equation~\eqref{eq:rho}, as $q^2$ is proportional to the density.\footnote{By taking the logarithm, equation~\eqref{eq:q_isoth} can also be rewritten in terms of $s$ or $c_0^2 s$, where $s=\log\rho$ is the logarithmic density fluctuation which is commonly used in the context of isothermal compressible turbulence.} Thus, we consistently recover the reduction of the system of three partial differential equations~\eqref{eq:rho}-\eqref{eq:rhoe} to the two equations of isothermal gas dynamics if the temperature is constant.

\subsection{Shell-to-shell transfer functions}
\label{sec:TransferFunct}

Following the notation introduced in \cite{Grete2017a}, the spectral energy transfer (for $\T{} > 0$) from a wave number shell $\shellQ$ of energy reservoir $\mathrm{X}$ to shell $\shellK$ of energy reservoir  $\mathrm{Y}$ is generically denoted by $\T{XY}(\shellQ,\shellK)$. The definition of a shell $\shellK$ (or $\shellQ$) is generally arbitrary from a formal point of view. We are using shells with equal distance in log space as describe in Sec.~\ref{sec:numericalTransfer}. Here, we use the symbols U and S for the energy reservoirs associated with the shell filtered $\V{w}$ and $q$ variables, respectively. For example,
\begin{align}
\label{eq:DefShell}
\shell{\V{w}}{K}\bra{\V{x}} = 
\int_\shellK \FT{\V{w}} \bra{\V{k}} \Eulere^{\imagi \V{k}\cdot\V{x}} 
\mathrm{d}\V{k}
\;,
\end{align}
where the integral is over all wavevectors within shell $\shellK$. The sum over all shells yields the field value in real space:
\begin{align}
\label{eq:fullShell}
\V{w}\bra{\V{x}} = \sum_\shellK\shell{\V{w}}{K}\bra{\V{x}} \;.
\end{align}
In contrast to momentum, the spatial average $\langle\V{w}\rangle$ does not vanish if the forcing is statistically isotropic. For this reason, modes at wavenumber zero in Fourier space (which correspond to mean values in physical space) are discarded for the computation of energy spectra and transfers. For a detailed discussion of shell averages, we refer the reader to \cite{Mininni2011,Grete2017a}.

The proper turbulent energy cascade is essentially given by energy transfers within the kinetic energy reservoir,
corresponding to the first two terms on the right hand side of the kinetic energy equation~\eqref{eq:KinEnFT}:
\begin{align}
\T{UU}(\shellQ,\shellK) = \T{UUa}(\shellQ,\shellK) + \T{UUc}(\shellQ,\shellK)\;,
\end{align}
where the advective and compressive components are defined by
\begin{align}
\label{eq:TransfKinAdv}
\T{UUa}(\shellQ,\shellK) &= -\int \Vshell{w}{K} \cdot \bra{\V{u} \cdot \nabla} \Vshell{w}{Q}\mathrm{d}\V{x}\;,  \\
\label{eq:TransfKinComp}
\T{UUc}(\shellQ,\shellK) &= -\frac{1}{2}\int \Vshell{w}{K} \cdot \Vshell{w}{Q} \nabla \cdot \V{u}\,\mathrm{d}\V{x} \;.
\end{align}
For the incompressible case ($d \equiv \nabla \cdot \V{u} = 0$), $\T{UUc}(\shellQ,\shellK)$ vanishes.

From the internal energy equation~\eqref{eq:IntEnFT}, analogous formulas are
obtained for the internal energy reservoir: 
\begin{align}
\T{SS}(\shellQ,\shellK) = \T{SSa}(\shellQ,\shellK) + \T{SSc}(\shellQ,\shellK)\;,
\end{align}
where
\begin{align}
\label{eq:TransfIntAdv}
\T{SSa}(\shellQ,\shellK) &= -\frac{2}{\gamma(\gamma-1)} \int 
             \shell{q}{K} \cdot \bra{\V{u} \cdot \nabla} \shell{q}{Q}\dd\V{x}\;,  \\
\label{eq:TransfIntComp}
\T{SSc}(\shellQ,\shellK) &= -\frac{1}{\gamma(\gamma-1)} \int  
             \shell{q}{K} \cdot \shell{q}{Q} \nabla \cdot \V{u}\,\dd\V{x} \;.
\end{align} 

Both $\T{UU}$ and $\T{SS}$ satisfy antisymmetry with themselves. Generally, antisymmetry states that energy gained in a shell  $\shellK$ of budget Y from a shell $\shellQ$ of  budget X must be equal to the energy that is lost from a shell $\shellQ$ of budget X to a shell $\shellK$ of budget Y. The antisymmetry property can be formally expressed as
\begin{align}
\T{XY}(\shellQ,\shellK) = - \T{YX}(\shellK,\shellQ) \;.
\end{align}
However, individual components such as $\T{SSa}$ and $\T{SSc}$ are not antisymmetric.

In \cite{Grete2017a} transfers from the internal to the kinetic
energy reservoir are defined by
\begin{align}
\label{eq:TransfPress}
	\T{PU}(\shellQ,\shellK) = - \int \frac{1}{\sqrt{\rho}} \Vshell{w}{K}\cdot \nabla \shell{p}{Q}
	\mathrm{d}\V{x} \;.
\end{align}
Since the total energy~(\ref{eq:TotEnDef}) is a conserved quantity (excluding sources and dissipation), exchanges between kinetic and internal energy integrated over all shells must cancel each other out. This can be shown by applying Parseval's theorem and integration by parts:
\begin{align*}
\sum_\shellK\sum_\shellQ &\T{PU}(\shellQ,\shellK) 
= -\int u_i (\pd_i p)\,\dd\V{x} \\
&= -\underbrace{\int \pd_i(u_i p)\,\dd\V{x}}_{=0\ \mbox{\scriptsize(periodic BC)}} + \int p \pd_i u_i\dd\V{x} 
= \int\frac{q^2}{\gamma}(\pd_i u_i)\,\dd\V{x} \\
&= -(\gamma-1)\sum_\shellK\sum_\shellQ \T{SSc}(\shellQ,\shellK)\;,
\end{align*}
where $(\gamma-1)\T{SSc}(\shellQ,\shellK)$ corresponds to the third term in equation~\eqref{eq:IntEnFT}.

While $\T{PU}(\shellQ,\shellK)$ and $\T{SSc}(\shellQ,\shellK)$ follow directly form equations~(\ref{eq:KinEnFT}) and~(\ref{eq:IntEnFT}),
antisymmetric transfer functions from internal to kinetic energy and vice versa can be defined in terms of the variable $q$ as
\begin{align}
\label{eq:TransfIntKin}
\T{SU}(\shellQ,\shellK) &= - \frac{1}{\gamma} \int
\frac{\Vshell{w}{K}}{\sqrt{\rho}} \cdot \nabla q \shell{q}{Q} 
\mathrm{d}\V{x} \;,\\
\label{eq:TransfKinInt}
\T{US}(\shellQ,\shellK) &= - \frac{1}{\gamma} \int 
\shell{q}{K} q \nabla \cdot \bra{\frac{\Vshell{w}{Q}}{\sqrt{\rho}}}
\mathrm{d}\V{x} \;.
\end{align}
Owing to their antisymmetry, $\T{SU}(\shellQ,\shellK)$ and $\T{US}(\shellQ,\shellK)$ cancel each other for individual shell-to-shell transfers between kinetic and internal energy. In contrast, $\T{PU}$ and $(\gamma-1)\T{SSc}$ compensate each other only in the global energy budget (sum over all $\shellQ$ and $\shellK)$).

In our approach, information about the coupling of the density field to other quantities is implicitly contained in the transfer functions by using the density weighted variables $\V{w}$ and $q$. Thus, we consider a simplified case of triadic interactions instead of quartic interactions arising from variable density.

\subsection{Shell energy equations}
\label{sec:ShellEn}

The rate of change of the kinetic and internal energy contained in a shell at wavenumber $K$ is given by the transfer functions defined in Section~\ref{sec:TransferFunct}:
\begin{align}
\label{eq:RateEkin_shell}
\pd_t \Ekin^\shellK &= 
\T{UUa}^\shellK + \T{UUc}^\shellK + \T{SU}^\shellK + \mathcal{F}^\shellK\\
\label{eq:RateEint_shell}
\pd_t \Eint^{\shellK} &= 
\T{SSa}^\shellK + \T{SSc}^\shellK + \T{US}^\shellK + \mathcal{S}^\shellK
\end{align}
The shell energies $\Ekin^\shellK$ and $\Eint^\shellK$ are obtained by integrating the energy densities over all wavenumbers belonging to shell $\shellK$ \cite{Grete2017a}:
\begin{align}
\label{eq:Ekin_shell}
\Ekin^\shellK &= \int_\shellK \Ekin (\V{k}) \dd\V{k}
=\int_\shellK \frac{1}{2} \FT{w}_i \FT{w}_i^* \dd\V{k}\\
\label{eq:Eint_shell}
\Eint^\shellK &= \int_\shellK \Eint (\V{k}) \dd\V{k}
=\int_\shellK \frac{1}{2} \FT{q} \FT{q}^* \dd\V{k}
\end{align}
Apart from a factor $\Delta\shellK$, which is the shell thickness, the shell energies can be understood as discrete energy spectra specifying the mean energy density at wavenumbers around $\shellK$ (see Section~\ref{sec:numericalSpect}).

All transfer terms $\T{XY}^\shellK$ in equations~\eqref{eq:RateEkin_shell} and~\eqref{eq:RateEint_shell} are defined as
\begin{align}
\label{eq:TransfShellK}
\T{XY}^\shellK = \sum_\shellQ\T{XY}(\shellQ,\shellK).
\end{align}
Generally, they specify the total rate of energy exchange between reservoirs X and Y through transfer from all shells $\shellQ$ to shell $\shellK$. In other words, they provide a measure of how energy within a particular shell $\shellK$ evolves via different interactions. Unless stated otherwise, energy transfer refers to $\T{XY}^\shellK$, not individual shell-to-shell transfer, in the following. The sources $\mathcal{F}^\shellK$ and $\mathcal{S}^\shellK$ stem from the forcing and cooling/heating terms, respectively. These terms are not considered in more detail here, as we are mainly interested in the dynamics of the turbulent cascade. One should bear in mind, however, that the sources may change the energy content of shells on top of the energy transfer mediated by nonlinear interactions (see also \cite{Grete2017a}).

For a realistic model of turbulent flow, a dissipation mechanism is essential. The numerical truncation errors of finite volume methods employed in many codes for compressible fluid dynamics usually mimic viscous dissipation, an approach which is sometimes called implicit large eddy simulation (ILES). Formally this can be written as
\begin{align}
\pd_t \Ekin^\shellK = 
\T{UUa}^\shellK + \T{UUc}^\shellK + \T{SU}^\shellK + \mathcal{F}^\shellK - \mathcal{D}^\shellK\;,
\end{align}
where $\mathcal{D}^\shellK$ is not explicitly known if dissipation is of numerical origin (this applies to the simulations presented in this article). In those cases where the compressible Navier-Stokes equations are solved or an explicit subgrid-scale model is applied, $\mathcal{D}^\shellK$ can be computed explicitly from the expression for the rate of energy dissipation $\epsilon$ \cite{Schmidt2015}. In the absence of sources, finite volume methods usually conserve total energy to machine precision. As a result, dissipation of kinetic energy is compensated by increasing internal energy:
\begin{align}
\pd_t \Eint^\shellK = 
\T{SSa}^\shellK + \T{SSc}^\shellK + \T{US}^\shellK + \mathcal{S}^\shellK + \mathcal{D}^\shellK\;,
\end{align}

The flow becomes statistically stationary once energy injection due to forcing is balanced by numerical dissipation and transfers to internal energy, i.e.
\begin{align}
\sum_\shellK \Big(\T{SU}^\shellK + \mathcal{F}^\shellK - \mathcal{D}^\shellK\Big) \sim 0\;.
\end{align}
Isothermal turbulence with a steady Mach number can only be produced if the excess heat produced by transfer from kinetic energy and dissipation is in turn removed by cooling, i.e.
\begin{align}
\sum_\shellK \Big(\T{US}^\shellK + \mathcal{S}^\shellK + \mathcal{D}^\shellK\Big) \sim 0\;.
\end{align}

In the case of incompressible turbulence, the inertial range is defined by $\T{UUa}^\shellK \simeq 0$ for shells with negligible forcing and dissipation.
Equivalently, the compresible kinetic energy transfer function must vanish in an ideal inertial range:
\begin{align}
\label{eq:TransfKinInertRange}
\pd_t\Ekin^\shellK \simeq\, \T{UUa}^\shellK + \T{UUc}^\shellK &\simeq 0\;.
\end{align}
However, the condition for an inertial range can be relaxed if we only require that the total energy is approximately constant in shells for which forcing, dissipation, and cooling are negligible:
\begin{align}
\label{eq:TransfInertRange}
\pd_t (\Ekin^\shellK + \Eint^\shellK) \simeq 
\T{tot}^\shellK \simeq 0\;,
\end{align}
where
\begin{equation}
\T{tot}^\shellK = \T{UU}^\shellK + \T{SU}^\shellK 
  + \T{SS}^\shellK + \T{US}^\shellK \simeq 0\;.
\end{equation}
Here, the transfers across energy budgets, $\T{SU}^\shellK$ and $\T{US}^\shellK$, do not necessarily cancel out for a given shell $\shellK$ (the antisymmetry relation holds in diagonal directions in the $\shellQ\shellK$-plane, while the shell equations are obtained by integration in $\shellQ$-direction for constant $\shellK$; see Fig.~\ref{fig:sketch}). Of course, it is not clear that any of the above conditions are met for compressible turbulence with external forcing and cooling. Regardless of whether an acceleration field $\V{a}$ or a force field $\V{f}=\rho\V{a}$ is used to inject energy in the simulation, the source term $u_i f_i = (\rho u_i)a_i$ in the kinetic energy equation always couples small and large scale modes, even for smooth acceleration (or force) fields (see \cite{Grete2017a} for a more detailed discussion). Moreover, heating and cooling can affect any wavenumber. In Section~\ref{sec:numericalTransfer} we will address the question of whether shells exist in which the energy transfers approximately sum up to zero (compared to their maximal values).

\subsection{Cross-scale energy fluxes}
\label{sec:DefCross}

By integrating transfer functions over all wavenumbers $Q \leq k$ and $K > k$ for a given wavenumber $k$, the cross-scale energy fluxes are obtained \cite{Debliquy2005,Grete2017a}:
\begin{align}
\label{eq:CrossFluxDef}
\Pi^{\mathrm{X}^<}_{\mathrm{Y}^>} (k) = \sum_{\shellQ \leq k} \sum_{\shellK > k} \T{XY}(\shellQ,\shellK),
\end{align}
The energy flux specifies the total rate of energy exchange between energy in reservoir $\mathrm{X}$ at wavenumbers smaller than $k$ and energy in reservoir $\mathrm{Y}$ at wavenumbers larger than $k$ (i.e.\ smaller length scales). This can be interpreted as scale-by-scale separation of the energy budget via low-pass and high-pass filters (see Section 2.4 in \cite{Frisch1995}). Summation over shells $\shellQ \leq k$ corresponds to length scales larger than $2\pi/ k$ and over shells $\shellK > k$ to smaller length scales. The energy flux is positive if there is a net transport of energy from larger to smaller scales.

\begin{figure}[tbp]
\centering
\includegraphics[width=0.4\textwidth]{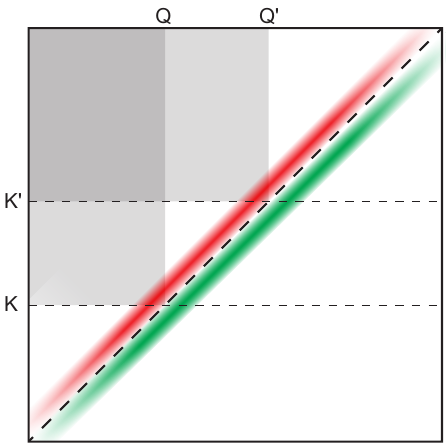}
\caption{Schematic view of energy transfer in the $\shellQ$-$\shellK$ plane for a local energy cascade. Shell-to-shell transfers are shown in color, with red and green indicating positive and negative transfers, respectively. The long dashed line for which $\shellQ=\shellK$ is the median line. The transfer terms on the right hand side of the shell energy equations~\eqref{eq:Ekin_shell} and ~\eqref{eq:Eint_shell} are given by summation over all $\shellQ$ for given $\shellK$ (equation~\ref{eq:TransfShellK}), corresponding to the horizontal dashed lines. Cross-scale fluxes are obtained by summing over rectangular regions $\shellQ\le k$ and $\shellK> k$, where $k$ is an arbitrary but fixed wavenumber. If the local energy transfers vanish sufficiently fast with distance from the median line (locality of energy transfer), the cross-scale fluxes for the gray-shaded regions are approximately constant.}
\label{fig:sketch}
\end{figure}

For incompressible turbulence, the constant kinetic energy flux $\Pi^{\mathrm{U}^<}_{\mathrm{U}^>}(k)$ is the defining property of the inertial subrange of wavenumbers. In general, the constancy of any cross-scale flux defined by equation~\eqref{eq:CrossFluxDef} requires two properties of the underlying shell-to-shell energy transfers in the inertial range. Firstly, $\T{XY}(\shellQ,\shellK)$ must be invariant under a diagonal wavenumber shift $(\shellQ,\shellK)\rightarrow(\shellQ^\prime,\shellK^\prime)$, where $\shellQ^\prime-\shellQ=\shellK^\prime-\shellK$. Secondly, $\T{XY}(\shellQ,\shellK)$ has to vanish sufficiently fast away from the median line $\shellQ = \shellK$ in the $\shellQ$-$\shellK$ plane. In other words, non-local transfers between distant shells must be negligibly small. This can be seen geometrically by observing that the index range in equation~(\ref{eq:CrossFluxDef}) covers a rectangular region to the left ($\shellQ \leq k$) and above ($\shellK > k$) the median line in the $\shellQ$-$\shellK$ plane, with the lower right corner touching the line at wavenumber $k$, as illustrated in Fig.~\ref{fig:sketch}. The numerical data in \cite{Grete2017a} demonstrate that invariance and locality of $\T{UU}(\shellQ,\shellK)$ is approximately satisfied for a limited range of wavenumbers in compressible MHD turbulence. However, deviations can be seen for transfers between kinetic and magnetic energy budgets. 

\section{Numerical simulations}
\label{sec:numericalData}

To run numerical simulations of statistically stationary and isotropic turbulence in a box with periodic boundary conditions, we implemented the stochastic forcing method from \cite{Schmidt2006,Schmidt2009} into the astrophysical fluid and N-body dynamics code Nyx \cite{Almgren2013}. A dimensionally unsplit Godunov method with full corner coupling and piecewise linear reconstruction is applied to solve equations of gas dynamics \cite{Colella1990,Miller2002}. This solver is particularly suitable for turbulence because it avoids spurious instabilities stemming from directional splitting \cite{Almgren2010}. For further details about the code and the postprocessing, see Appendix~\ref{sec:compute}. 

We follow the reasoning of \cite{Wagner2012,Aluie2013} in maintaining compressible turbulence by a large-scale acceleration rather than a body force. The basic idea of stochastic forcing is to compose a random acceleration field $\V{a}(\V{x},t)$ that varies smoothly not only in space, but also in time. It was recently demonstrated that the autocorrelation time of this field has a significant impact on statistical properties of forced turbulence \cite{Grete2018}. We set the autorcorrelation time equal to the dynamical time scale $T$, which is in turn determined by the amplitude and length scale of the forcing. In the following, we use the basic parameters $L$, $V$, and $\zeta$ introduced in \cite{Schmidt2006,Schmidt2009} to specify the properties of the forcing (see Table~\ref{tab:simulations}). The length scale $L$ is defined such that $2\pi/L$ is the wavenumber at which the spectrum of the forcing has its peak (we choose $L$ to be half the box size). By scaling the forcing amplitude with $V^2/L$, the large-scale velocity fluctuations induced by the forcing become comparable to $V$ in the statistically stationary regime (for this reason, $V$ is called the characteristic velocity of the flow). The dynamical time scale is given by $T=L/V$. The relative strength of the solenodial (divergence-free) and compressive (rotation-free) components of the forcing can be chosen by means of a Helmholtz decomposition in spectral space with weight coefficients $\zeta$ and $1-\zeta$, respectively. The two limiting cases of purely solenoidal ($\zeta=1$) and compressive ($\zeta=0$) forcing were investigated in \cite{Schmidt2008,Federrath2010,Federrath2013}.

\begin{table}[tbp]
        \begin{tabular}{lllcc}
                \toprule
                $V$ & $\zeta$ & $\alpha$ & $\langle w^2\rangle^{1/2}$ & 
                $\langle\mathrm{Ma}^2\rangle^{1/2}$ \\
\colrule
0.25 & 1 & 1000 & 0.183 & 0.497 \\
\colrule
0.5 & 1 & 1000 & 0.338 & 0.948 \\
0.5 & 2/3 & 1000 & 0.329 & 0.920 \\
\colrule
1.0 & 1 & 1000 & 0.596 & 1.715 \\
1.0 & 2/3 & 1000 & 0.587  & 1.691 \\
1.0 & 2/3 & 100 & 0.592 & 1.694 \\
1.0 & 2/3 & 10 & 0.606 & 1.688 \\
1.0 & 2/3 & 1 & n/a & n/a \\
1.0 & 1/2 & 1000 & 0.554 & 1.585 \\
\colrule
2.0 & 2/3 & 1000 & 1.114 & 3.189 \\
2.0 & 1/2 & 1000 & 1.044 & 3.018 \\
2.0 & 1/4 & 1000 & 0.830 & 2.452 \\
\colrule
4.0 & 2/3 & 1000 & 2.160 & 6.090 \\
4.0 & 1/2 & 1000 & 2.066 & 5.860 \\
4.0 & 1/4 & 1000 & 1.705 & 5.114 \\
4.0 & 1/4 & 10 & n/a & n/a \\
4.0 & 1/8 & 1000 & 1.596 & 4.929 \\
\botrule
        \end{tabular}
        \caption{Overview of simulation parameters (integral velocity scale $V$, solenoidal weight parameter $\zeta$, cooling cofficient $\alpha$). Also listed are the time-averaged RMS mass-weighted velocity $w$ and RMS Mach number defined by equation~\eqref{eq:rms_mach} in the statistically stationary regime.}
        \label{tab:simulations}
\end{table}

\subsection{Existence of a statistically stationary regime}
\label{sec:stat}

The initial mass and energy densities in code units are $\rho_0=1$ and energy $E_{\rm int,0}\approx 0.25$, respectively, for all simulations. Since turbulence is scale-free, the absolute value of the energy is of no particular significance and we will consider mostly dimensionless relative quantities. 

\begin{figure*}[!htbp]
\centering
\includegraphics[width=0.48\textwidth]{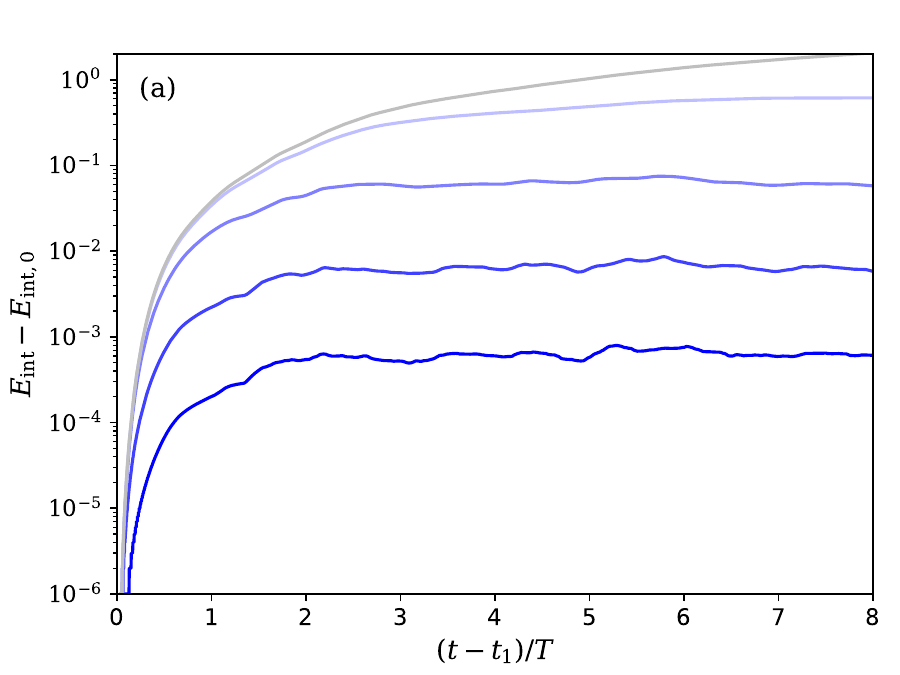}
\includegraphics[width=0.48\textwidth]{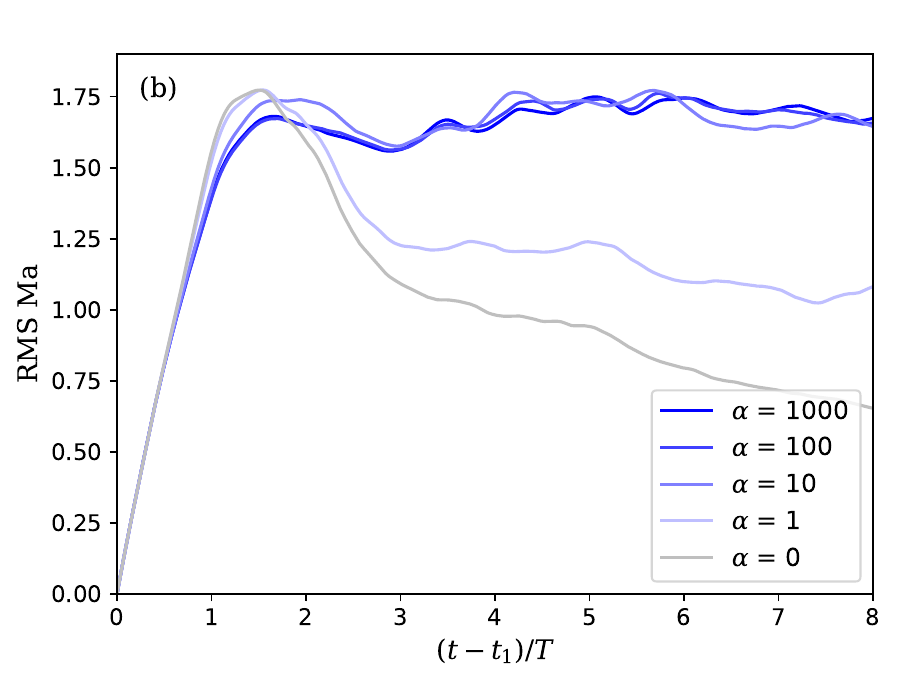}
\caption{Time evolution of the mean internal energy minus initial energy (code units) and RMS Mach number for different cooling coefficients $\alpha$ in simulations with forcing parameters $V=1.0$ and $\zeta=2/3$ (see also ~\ref{tab:simulations}).}
\label{fig:mean_alpha}
\end{figure*}

\begin{figure*}[tbp]
\centering
\includegraphics[width=0.45\textwidth]{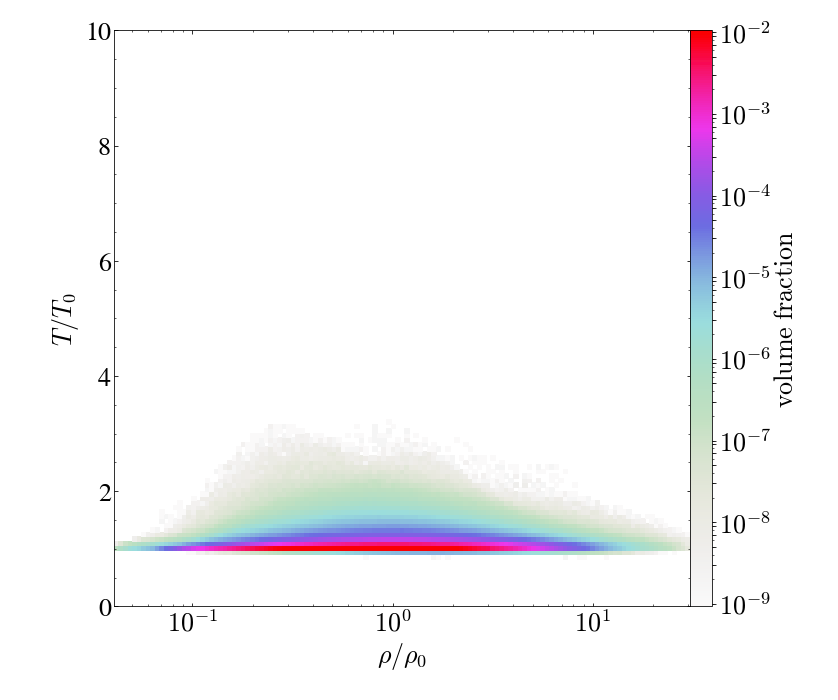}
\includegraphics[width=0.45\textwidth]{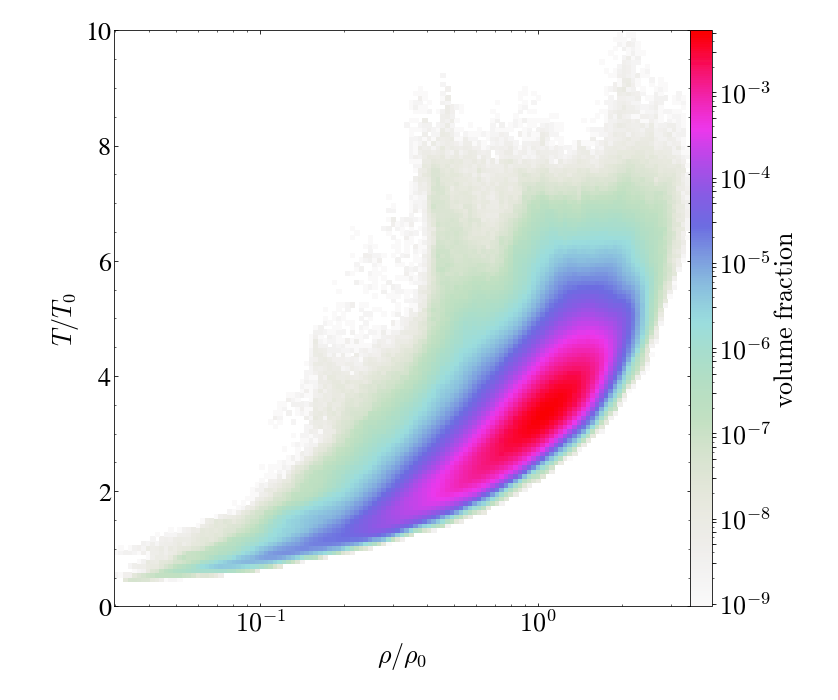}
\caption{Phase diagrams for cooling coefficients $\alpha=1000$ (left) and $1$ (right) at the end of simulations with $V=1.0$ and $\zeta=2/3$. The temperature $T$ and density $\rho$ are normalized to their initial values.}
\label{fig:phase}
\end{figure*}

In contrast to most simulations of isothermal turbulence in the literature, we do not enforce an isothermal equation of state ($p\propto\rho$) or use an artificially increased internal energy $\Eint=\rho e=p/(\gamma-1)$, where $\gamma-1$ is a small fraction of unity. In our simulations, $\gamma=5/3$ and cooling of the gas is mimicked by the term
\begin{equation}
    \Lambda_{\rm HC}=-\alpha\rho(e-e_0)
    \label{eq:cooling}
\end{equation}
in equation~(\ref{eq:rhoe}), where $e_0$ is the initial gas energy per unit mass. Excess heat $e-e_0$ is removed at a rate given by the cooling coefficient $\alpha$ (if the energy difference is negative, the gas is heated). In the following, we show that this simple model results in a steady state with approximately constant mean thermal energy if $\alpha$ is chosen sufficiently large. In this case, the Mach number of the turbulent flow can be adjusted by
changing the forcing parameter~$V$.

\begin{figure*}[!htbp]
\centering
\includegraphics[width=\textwidth]{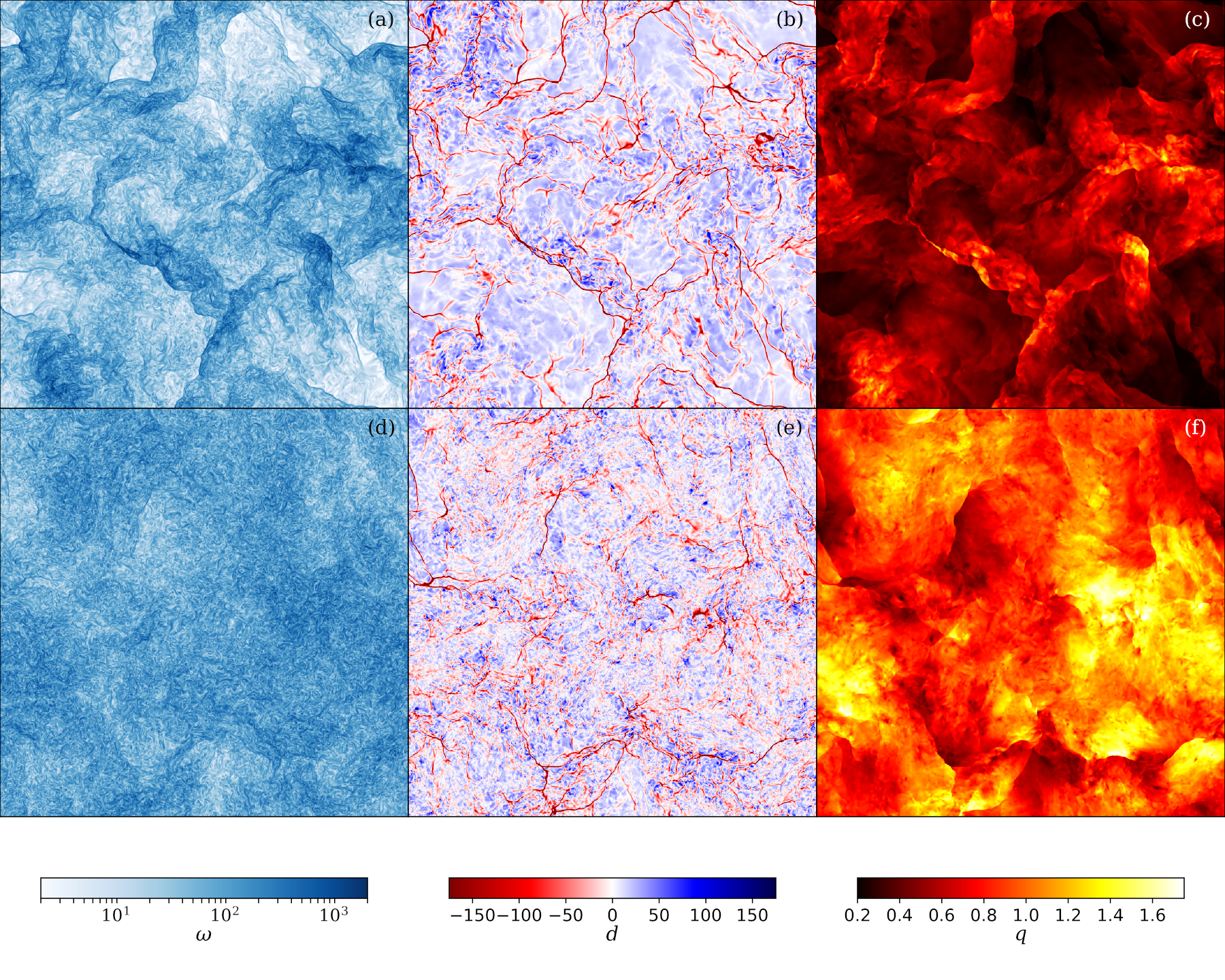}
\caption{Slices of the vorticity $\omega$ (a,d), divergence $d$ (b,e), and $q$ (c,f) defined by equation~(\ref{eq:q_def}) for cooling coefficients $\alpha=1000$ (a,b,c) and $1$ (d,e,f) at the end of simulations with $V=1.0$ and $\zeta=2/3$ (the corresponding phase diagrams are shown in Fig.~\ref{fig:phase}).}
\label{fig:slices}
\end{figure*}

First we consider the impact of the the cooling coefficient $\alpha$ in equation~(\ref{eq:cooling}) for the case $V=1.0$ and $\zeta=2/3$. As shown in Figure~\ref{fig:mean_alpha} (a),  the mean internal energy becomes approximately constant for $\alpha\gtrsim 10$ after an initial phase in which the gas is set into motion by the forcing and eventually becomes turbulent. The root mean square (RMS) Mach number 
\begin{equation}
    \label{eq:rms_mach}
    \langle\mathrm{Ma}^2\rangle^{1/2}=
    \langle u^2/c_{\rm s}^2\rangle^{1/2}=\langle w^2/q^2\rangle^{1/2}
\end{equation}
saturates at about $1.7$ (b; see also Table~\ref{tab:simulations} for time averages). This indicates a steady state in which the rate of mechanical energy injection is balanced by the rate of dissipation into heat, which in turn is balanced by the net cooling rate of the system. 
In this case, the mean kinetic and thermal energies are roughly constant and the gas is quasi isothermal with larger fluctuation for lower $\alpha$. The distribution of the temperature fluctuations for $\alpha=1000$ is shown in the left phase diagram in Fig.~\ref{fig:phase}.
For the smallest cooling coefficient $\alpha=1.0$, one can see in Fig.~\ref{fig:phase} that the internal energy gradually increases and the RMS Mach number slowly drifts into the subsonic regime (the adiabatic case, $\alpha = 0$, is shown as gray line). This marks the transition to more or less adiabatic behaviour (see the phase diagram of a simulation with $\alpha = 1.0$ in the right panel of Fig.~\ref{fig:phase}). Figure~\ref{fig:slices} illustrates the turbulent flow for these two cases. Panels (a,d) show slices of the vorticity modulus $\omega=|\nabla\times\V{u}|$. In the nearly isothermal case (a,b,c), the vorticity appears to be more intermittent with pronounced front-like features which are associated with strongly negative divergence $d=\nabla \cdot \V{u}$ (b,e). This is indicative of shock fronts. Shocks are also reflected in the pronounced jumps of $q$ (c,f), which is approximately the square root of the mass density $\rho$ if the gas is nearly isothermal. In contrast, $q$ is generally larger, i.e.\ the gas is hotter, and jumps are less pronounced for $\alpha=1$ (d,e,f).

\begin{figure*}[!htbp]
\centering
\includegraphics[width=0.48\textwidth]{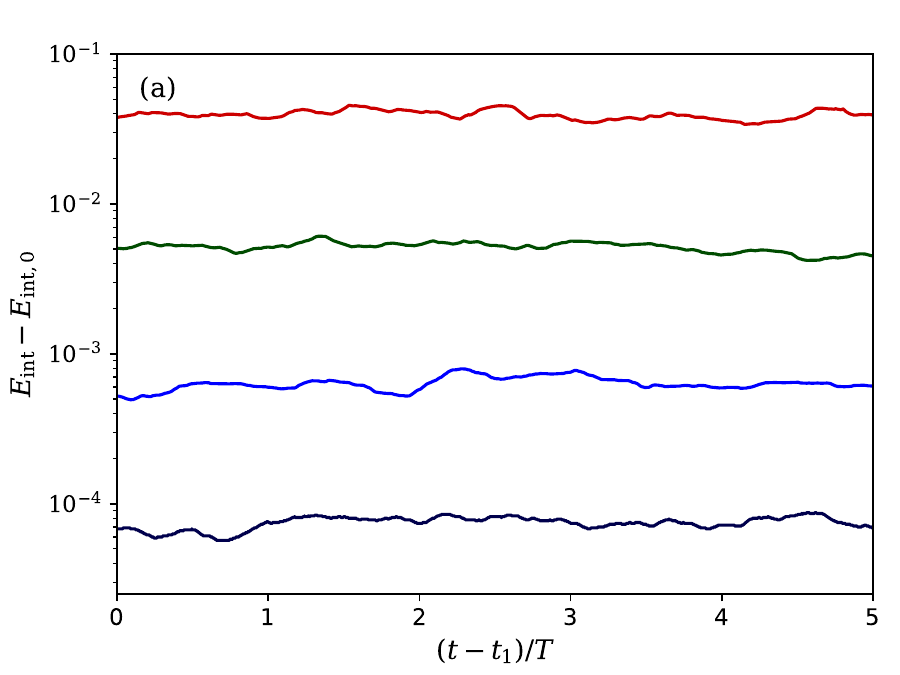}
\includegraphics[width=0.48\textwidth]{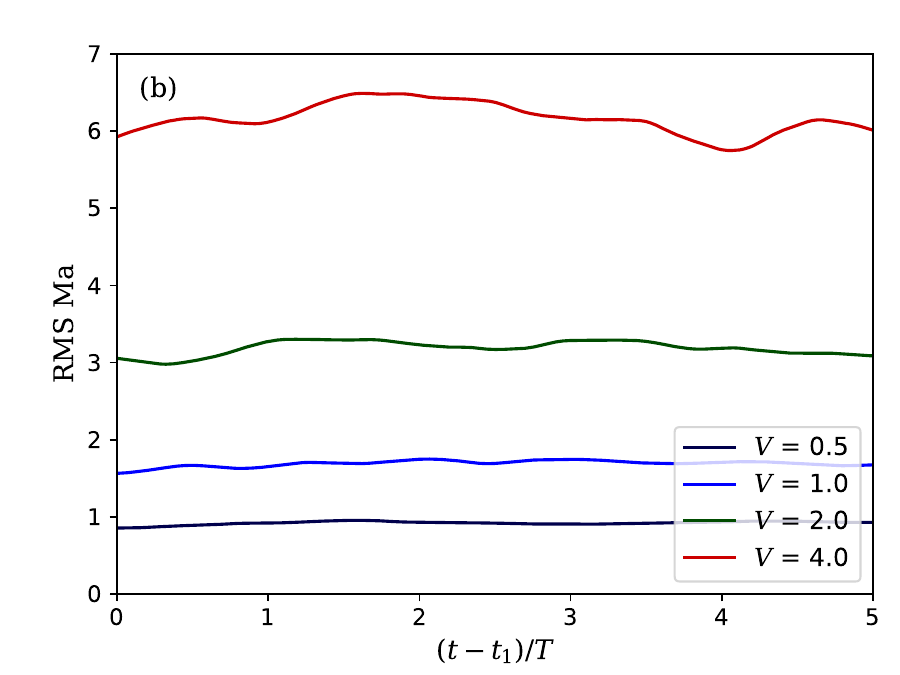}
\caption{Time evolution of the mean internal energy minus initial energy (a; in code units) and RMS Mach number (b) for different forcing amplitudes in combination with $\zeta=2/3$ and $\alpha=1000$ in the statistically stationary phase starting at $t=t_1$. }
\label{fig:mean_2to1}
\end{figure*}

In Fig.~\ref{fig:mean_2to1}, mean values for varying integral velocity scale $V$ are shown in the statistically stationary regime starting at $t=t_1$. We choose $t_1=3T$ for $V=1.0$ or lower (see Fig.~\ref{fig:mean_alpha}). For higher $V$, the forcing magnitude is ramped up in steps from 1.0 to 2.0 to 4.0 in order to avoid CFL violations triggered by very strong shocks and rarefactions in the initial phase (turbulence reduces these effects) and $t_1$ is accordingly adjusted. 
In all four cases, $\zeta=2/3$ and $\alpha=1000$. Figure~\ref{fig:mean_2to1} (a) shows that the deviation of the mean internal energy from $E_{\rm int,0}$, which is proportional to the net cooling rate (see equation~\ref{eq:cooling}), increases with the rate of energy injection by the forcing, while the mean energy remains at a nearly constant level. The RMS Mach numbers (b) indicate that $\alpha=1000$ is sufficient to maintain statistically stationary and nearly isothermal turbulence up to $V=4.0$. The resulting time-averaged RMS Mach numbers are listed in Table~\ref{tab:simulations}.

For supersonic isothermal turbulence, density fluctuations are expected to follow a log-normal distribution \cite{Passot1998,Kritsuk2007,Lemaster2008,Federrath2008, Schmidt2009,Konstandin2012}. In terms of the logarithmic density fluctuation $s=\log\rho$ (here it is assumed that the mean density in code units is unity), this distribution is defined by
\begin{equation}
    \label{eq:pdf}
    P(s)\dd s = \frac{1}{\sqrt{2\mathrm{\pi}\sigma_s}}\exp\left(-\frac{(s-s_0)^2}{2\sigma_s^2}\right)\dd s\;,
\end{equation}
where $s_0=-\sigma_s^2/2$ is implied by mass conservation (it should be noted that $s_0$ does not correspond to the mean density, which would imply $s_0=0$). The function $P(s)$ is a probability density function (PDF), i.e.\ the cumulative probability for finding $s$ in a given interval is obtained by integrating $P(s)$. 

\begin{figure*}[!htbp]
\centering
\includegraphics[width=0.48\textwidth]{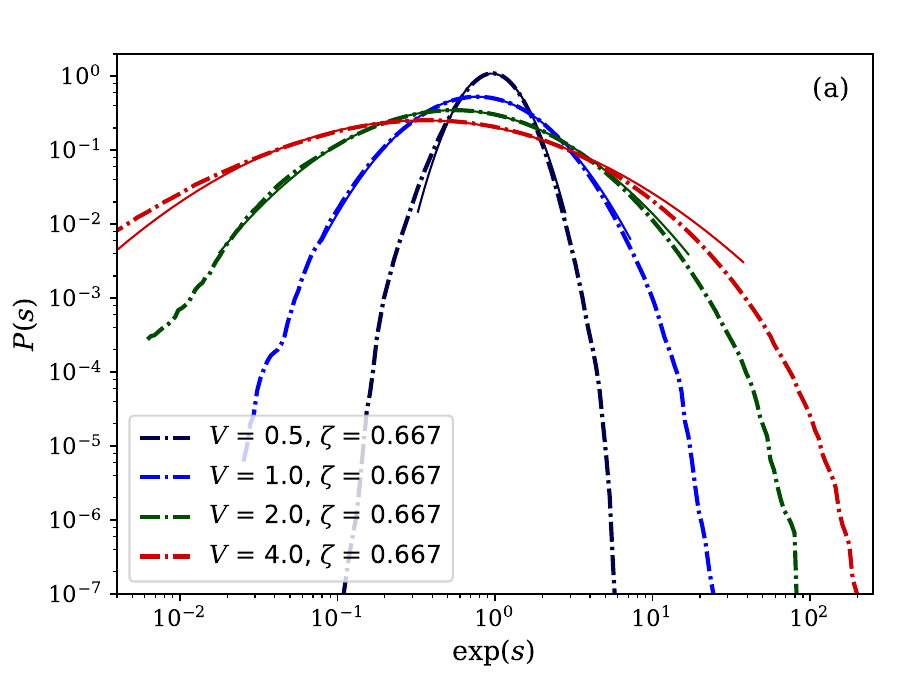}
\includegraphics[width=0.48\textwidth]{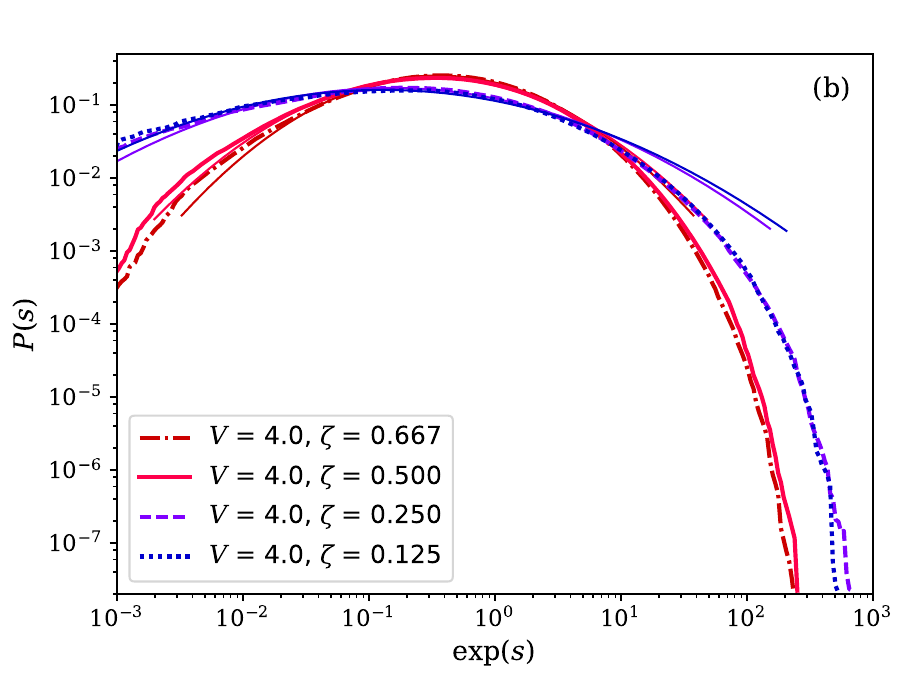}
\caption{Time-averaged PDFs of the logarithmic mass density $s=\log(\rho/\rho_0)$, where $\rho_0=1$, for different Mach numbers and fixed $\zeta=2/3$ (a) and for varying $\zeta$ (b). The thin solid lines indicate log-normal fits defined by \eqref{eq:pdf} in the range $-3\sigma_s \le s-s_0 \le 3\sigma_s$.}
\label{fig:pdf}
\end{figure*}

Samples of time-averaged PDFs from our simulations are plotted in Fig.~\ref{fig:pdf}. Plot (a) shows the dependence on the forcing magnitude for $\zeta=2/3$. The width of the PDF increases with the characteristic velocity of the flow, which simply reflects the stronger density contrast at higher Mach number \cite{Padoan1997,Konstandin2012}. This trend can also be seen by fitting equation~\eqref{eq:pdf} to the data. The resulting values of the standard deviation $\sigma_s$ are summarized in Table~\ref{tab:log_normal}. If the PDFs were exactly log-normal, $s_0$ would be fixed by $\sigma_s$. The deviation of $s_0 + \sigma_s^2/2$ from zero is thus a measure for the deviation of the PDF from log-normal shape. The deviations tend to increase with the Mach number.\footnote{Improved fits with intermittency correction were proposed in \cite{Hopkins2013a}} Figure~\ref{fig:pdf} (b) reveals that the PDF in the supersonic case depends strongly on $\zeta$. Depending on $\zeta$, the RMS velocity produced for a given forcing magnitude (characteristic velocity $V$) varies somewhat (see Table~\ref{tab:simulations}). Since RMS Mach numbers are slightly lower for strongly compressive forcing (as discussed in \cite{Schmidt2009}, this is related to stronger intermittency), PDFs should become narrower if the variations were solely due to Mach-dependent compressibility of the turbulent flow. The PDFs for $V=4.0$ are, on the contrary, substantially broader if the forcing is dominated by compressive modes ($\sigma\approx 2.5$ for $\zeta=1/8$ compared to $1.6$ for $\zeta=2/3$). The skewness also becomes more pronounced for low $\zeta$, as indicated by the large discrepancy between the log-normal fits and the data toward high overdensities ($\rho\gtrsim 10$). This effect of the forcing is extensively discussed in \cite{Federrath2008,Schmidt2009,Federrath2010,Konstandin2012,Konstandin2012b,Hopkins2013a,Federrath2013}. However, the log-normal density PDF is an idealization that applies to exactly isothermal gas with a self-similar hierarchy of density structures, which can exist only in the strongly supersonic case. Since the gas in our simulations is only approximately isothermal (particularly shock-heated gas does not cool instantaneously) and Mach numbers cover a range from below unity to about five, we find significant deviations in the far tails (beyond a few $\sigma_s$) in all cases. For our purpose it is sufficient that PDFs are close to log-normal, as a further indication of nearly isothermal turbulence. 

\begin{table}[tbp]
        \begin{tabular}{llcc}
                \toprule
                $V$ & $\zeta$ & $\sigma_s$ & $s_0 + \sigma_s^2/2$ \\
\colrule
0.25 & 1 & 0.117 & 0.013\\
\colrule
0.5 & 1 & 0.373 & 0.021\\
0.5 & 2/3 & 0.366 & 0.016\\
\colrule
1.0 & 1 & 0.735 & 0.023\\
1.0 & 2/3 & 0.754 & 0.017\\
1.0 & 1/2 & 0.778 & 0.035\\
\colrule
2.0 & 2/3 & 1.147 & 0.047\\
2.0 & 1/2 & 1.227 & 0.086\\
2.0 & 1/4 & 1.612 & 0.340\\
\colrule
4.0 & 2/3 &  1.574 & 0.188\\
4.0 & 1/2 & 1.689 & 0.247\\
4.0 & 1/4 & 2.326 & 0.798\\
4.0 & 1/8 & 2.474 & 1.006\\
\botrule
        \end{tabular}
        \caption{Standard deviations of two-parameter $(s_0, \sigma_s)$ log-normal fits to the PDFs of the mass density for $\alpha=1000$. For an exact log-normal distribution, mass conservation implies $s_0 + \sigma_s^2/2=0$. The actual values obtained from the fit functions are listed in the column on the very right.}
        \label{tab:log_normal}
\end{table}

\subsection{Transfer function}
\label{sec:numericalTransfer}

\begin{figure*}[!htbp]
\centering
\includegraphics[width=\textwidth]{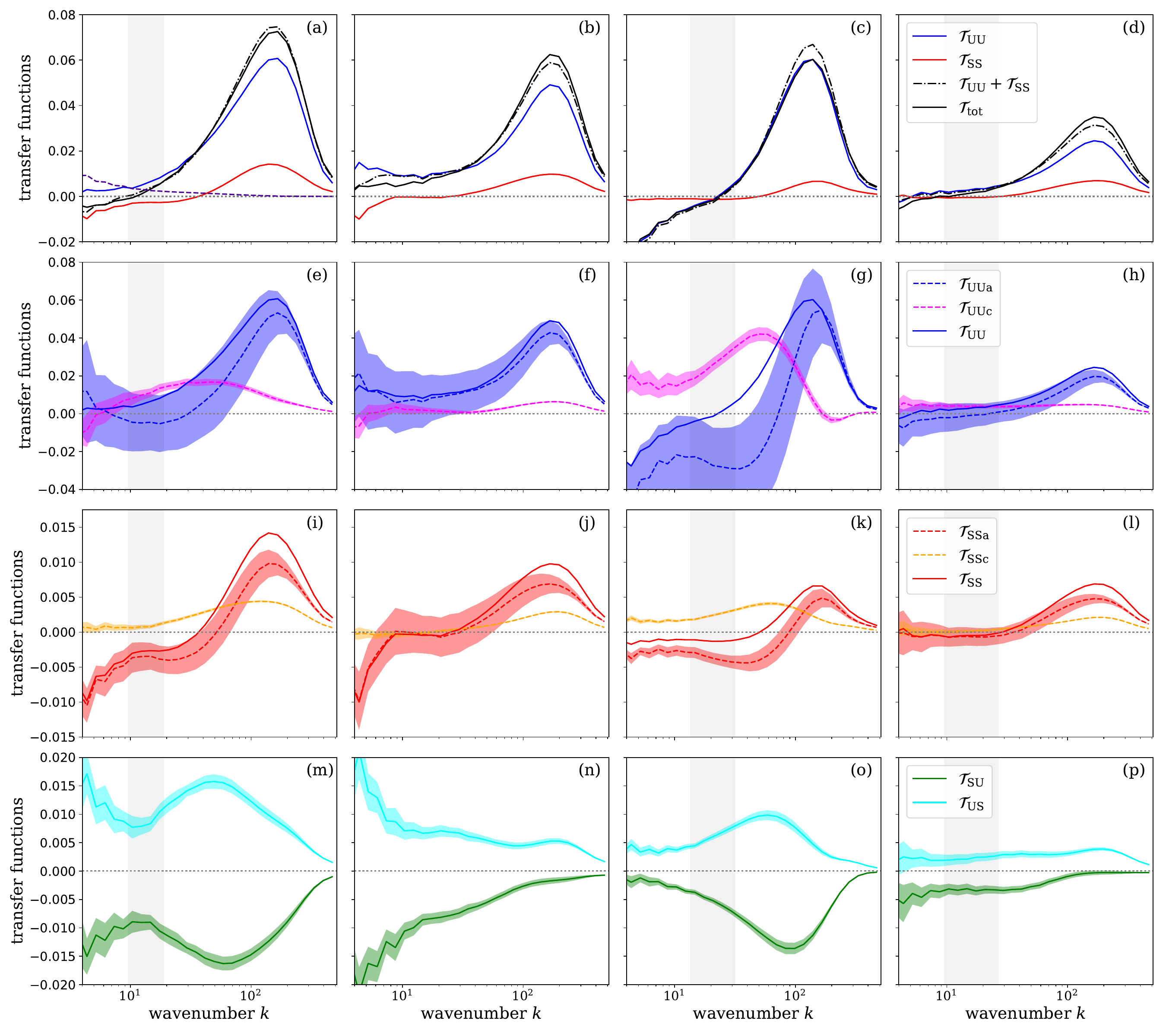}
\caption{Time-averaged transfers $\T{XY}^\shellK$ into shell $\shellK$ (see equation~\ref{eq:TransfShellK}) for energy reservoirs X and Y indicated by the legends in the right column. Forcing with $V=1.0$ and $\zeta=2/3$ was applied for the transfer functions in plots (a,e,i,m) and (b,f,j,n), while $V=4.0$ and $\zeta=1/4$ for (c,g,k,o) and (d,h,l,p). In both cases the cooling coefficient was varied: $\alpha=1000$ for (a,e,i,m) and (c,g,k,o), $\alpha=10$ for (b,f,j,n) and (d,h,l,p). See also Table~\ref{tab:simulations} for an overview. All transfers are scaled by $V^3$. The range between one standard deviation above and below the time average is filled (except for the top panels). The estimated inertial range on the basis of near zero transfer is indicated by gray shaded wavenumber ranges.  }
\label{fig:transf_alpha}
\end{figure*}

In Fig.~\ref{fig:transf_alpha}, we plot transfer functions $\T{UU}^\shellK$, $\T{SS}^\shellK$, $\T{SU}^\shellK$, and $\T{US}^\shellK$ defined as rate of change of energy in shell $\shellK$ for simulations representing moderate and high Mach numbers as well as strong and weak cooling (see Table~\ref{tab:simulations}). We adopt octave binning, i.e. $\shellK\,\equiv k \in [K/2,2K[$, as in \cite{Grete2017a}. In each case, the transfer functions are averaged over an interval of at least five dynamical timescales $T$ in steps of $0.2T$.

Figure~\ref{fig:transf_alpha} (a,b,c,d) shows the total energy transfers $\T{tot}^\shellK$ (solid black lines) and the sum of kinetic and internal energy transfers, $\T{UU}^\shellK+\T{SS}^\shellK$ (dot-dashed black lines) in comparison to $\T{UU}^\shellK$ and $\T{SS}^\shellK$. The energy-containing range at low wavenumbers is excluded here. It is rather obvious that there is no extended range of wavenumbers that could be interpreted as inertial range in the sense discussed in Section~\ref{sec:ShellEn}, i.e.\ neither $\T{tot}^\shellK$ or $\T{UU}^\shellK$ vanishes in between negative (energy-containing range) and positive (dissipation range) peaks. However, $\T{tot}^\shellK$ crosses zero in three out of four cases. As a working hypothesis, we adopt the criterion that an approximate inertial range requires $|\T{tot}^\shellK|$ to be less than $0.1$ times its peak value and $\T{tot}^\shellK$ to be negative or sufficiently close to zero ($0.01$ times the maximum) in at least one shell $\shellK \ge 10$ (at lower wavenumbers, the forcing significantly contributes to the energy transfer, which we checked for selected cases). The resulting wavenumber ranges are gray shaded in the plots and summarized for all simulations in Table~\ref{tab:powerlaw}. At best, there is a marginal inertial range for nearly isothermal turbulence ($\alpha=1000$) in Fig.~\ref{fig:transf_alpha} (a) and (c). For $\alpha=10$, there appears to be an imbalance of energy at intermediate scales and, as a result, no inertial range defined by our criterion exists in the case $V=1.0$, corresponding to Mach numbers around $1.7$ (b). In contrast, we find a relatively wide inertial range for $V=4.0$ (d) with forcing dominated by compressive modes ($\zeta=0.25$), even though the flow does not enter a statistically stationary regime in this case (the Mach number gradually decreases, as the cooling is not efficient enough). 
This shows that inertial-range behavior is limited by the interplay between the compressibility of the flow, the forcing, and thermodynamics.
This particularly applies to the forcing, as the coupling of the acceleration field to the velocity field through density variations in the compressible regime introduces significant net transfer of energy on the largest scale ($k \lesssim 10$), see also Fig. 7 in \citep{Grete2017a}.

Figure~\ref{fig:transf_alpha} (a,b,c,d) also shows that $\T{UU}^\shellK$ does not differ much from the total transfer $\T{tot}^\shellK$ in the strongly supersonic case. Consequently, we could base our estimate of the inertial range just as well on $\T{UU}^\shellK$ as on $\T{tot}^\shellK$. In other words, the kinetic energy is roughly an ideal invariant, as expressed by \eqref{eq:TransfKinInertRange}. This is simply a consequence of the internal energy being small in comparison to the kinetic energy in the limit of high Mach numbers. As observed for MHD turbulence in \cite{Grete2017a}, the advective and compressive components of $\T{UU}^\shellK$ (e,f,g,h) roughly cancel each other at intermediate wavenumbers (second row of plots), which results more or less in inertial-range behaviour of the kinetic energy. The peak between $k=100$ and $200$, which can be seen in all cases, indicates that energy received from lower wavenumbers (larger eddies) per unit time exceeds the energy that is drawn by higher wavenumbers (smaller eddies) in the forward cascade. This can be attributed to the increasing damping of eddies by numerical viscosity toward high wave numbers. Depending on the properties of the solver, this imbalance may result in a so-called bottleneck effect (see Section~\ref{sec:numericalSpect}).

For low $\alpha$ (i.e.\ the gas is close to adiabatic), the internal energy transfer $\T{SS}^\shellK$ is about zero at intermediate wavenumbers (panels (j,l) in Fig.~\ref{fig:transf_alpha}). This can be interpreted as an internal energy cascade in analogy to the usually considered cascade of turbulent kinetic energy, with shell-to-shell interactions between fluctuations of $q$ rather than $w$. There is also a similar peak at high wave numbers. The transfer of internal energy can be understood as turbulent mixing, which redistributes internal energy from larger to smaller scales. The mixing agent is the turbulent flow, corresponding to the advection operator $\V{u}\cdot\nabla$ occuring in equations~\eqref{eq:TransfKinAdv} and~\eqref{eq:TransfIntAdv} for kinetic and internal energy transfer, respectively (and the divergence operator in the expressions for the compressive components). In the case of nearly isothermal turbulence ($\alpha=1000$), the internal energy transfer in the estimated inertial range is negative (i,k). This reflects net energy losses due to cooling  on top of the redistribution of internal energy through turbulent mixing. Naturally, the component $\T{SSc}^\shellK$ becomes more significant in comparison to $\T{SSa}^\shellK$ with stronger shock compression at higher Mach numbers. 

Transfers from kinetic to internal energy reservoirs are of the same order of magnitude as $\T{SS}^\shellK$ (panels (m,n,o,p) in Fig.~\ref{fig:transf_alpha}). However, $\T{US}^\shellK$ and $\T{US}^\shellK$ nearly cancel each other, particularly at low to intermediate wavenumbers.
This is in agreement with results reported in \cite{Wang2018} using a filtering approach to
analyze sub- and transonic solenoidally forced turbulence simulation with explicit viscosity.
For this reason, $\T{UU}^\shellK+\T{SS}^\shellK$ can be used as a proxy for $\T{tot}^\shellK$ (solid black vs dot-dashed lines in top panels). We thus computed only $\T{UU}^\shellK$ and $\T{SS}^\shellK$ for most other simulations with $\alpha=1000$ to reduce the postprocessing time. Our results for $\T{US}^\shellK$ and $\T{US}^\shellK$ reveal a stronger exchange between the energy reservoirs for isothermal turbulence in the range of wavenumbers where the compressive components $\T{UUc}^\shellK$ and $\T{SSc}^\shellK$ are large. This is to be expected since $\T{UUc}^\shellK$, $\T{SSc}^\shellK$, and $\T{US}^\shellK$   are related to the divergence of the flow (see equations~\ref{eq:TransfKinComp}, ~\ref{eq:TransfIntComp}, and~\ref{eq:TransfKinInt}) and $\T{SU}^\shellK$ is the antisymmetric counterpart of $\T{US}^\shellK$.

\begin{figure*}[!htbp]
\centering
\includegraphics[width=\textwidth]{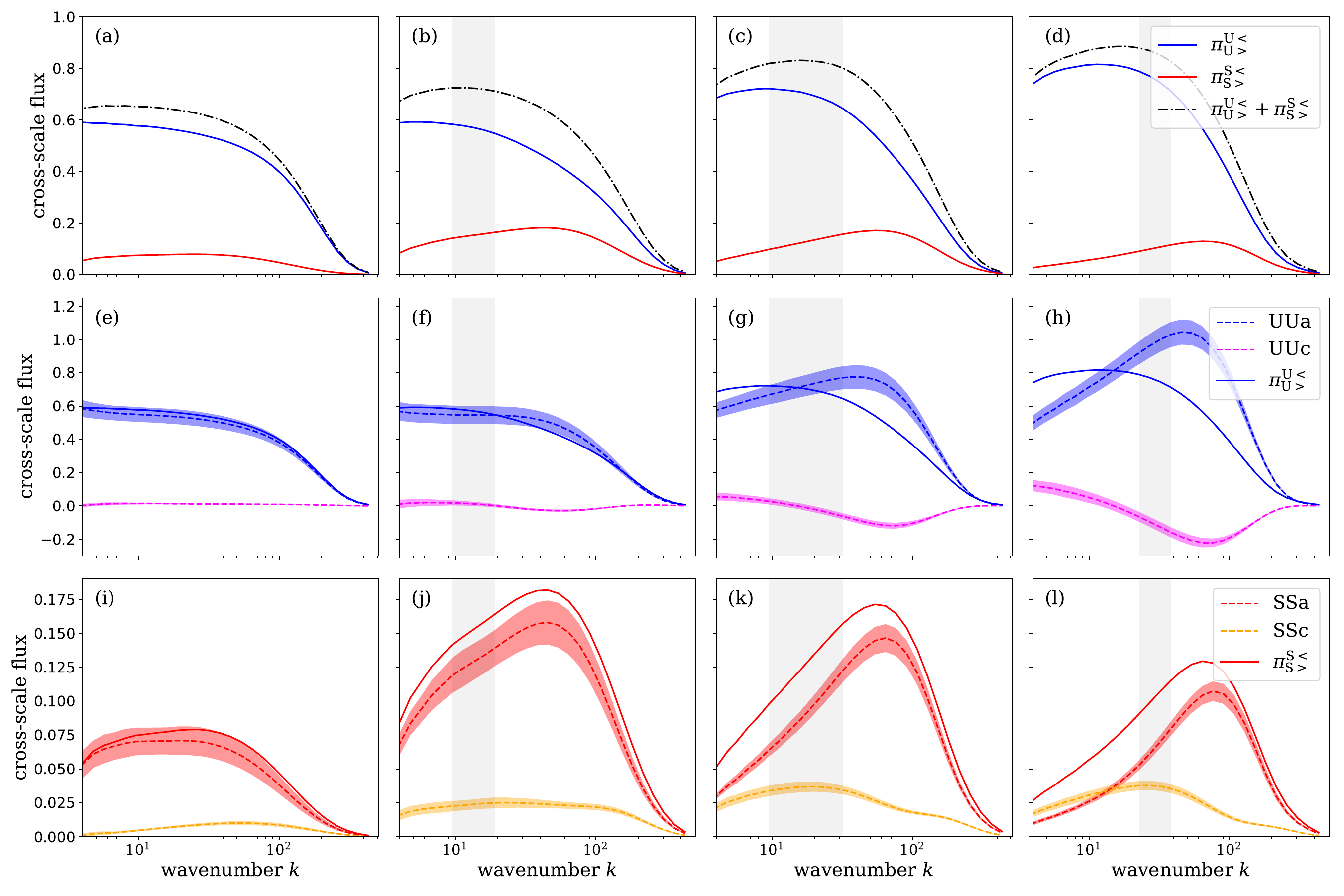}
\caption{Time-averaged cross-scale fluxes of kinetic and internal energy in nearly isothermal turbulence ($\alpha=1000$) for increasing forcing magnitude ($V=0.5$ for (a,e,i), $V=1.0$ for (b,f,j), $V=2.0$ for (c,g,k), and $V=4.0$ for (d,h,l)) and a constant mixing ratio $\zeta=2/3$. All fluxes are normalized by $V^3$.}
\label{fig:cross_scale_2to1}
\end{figure*}

\begin{figure*}[!htbp]
\centering
\includegraphics[width=\textwidth]{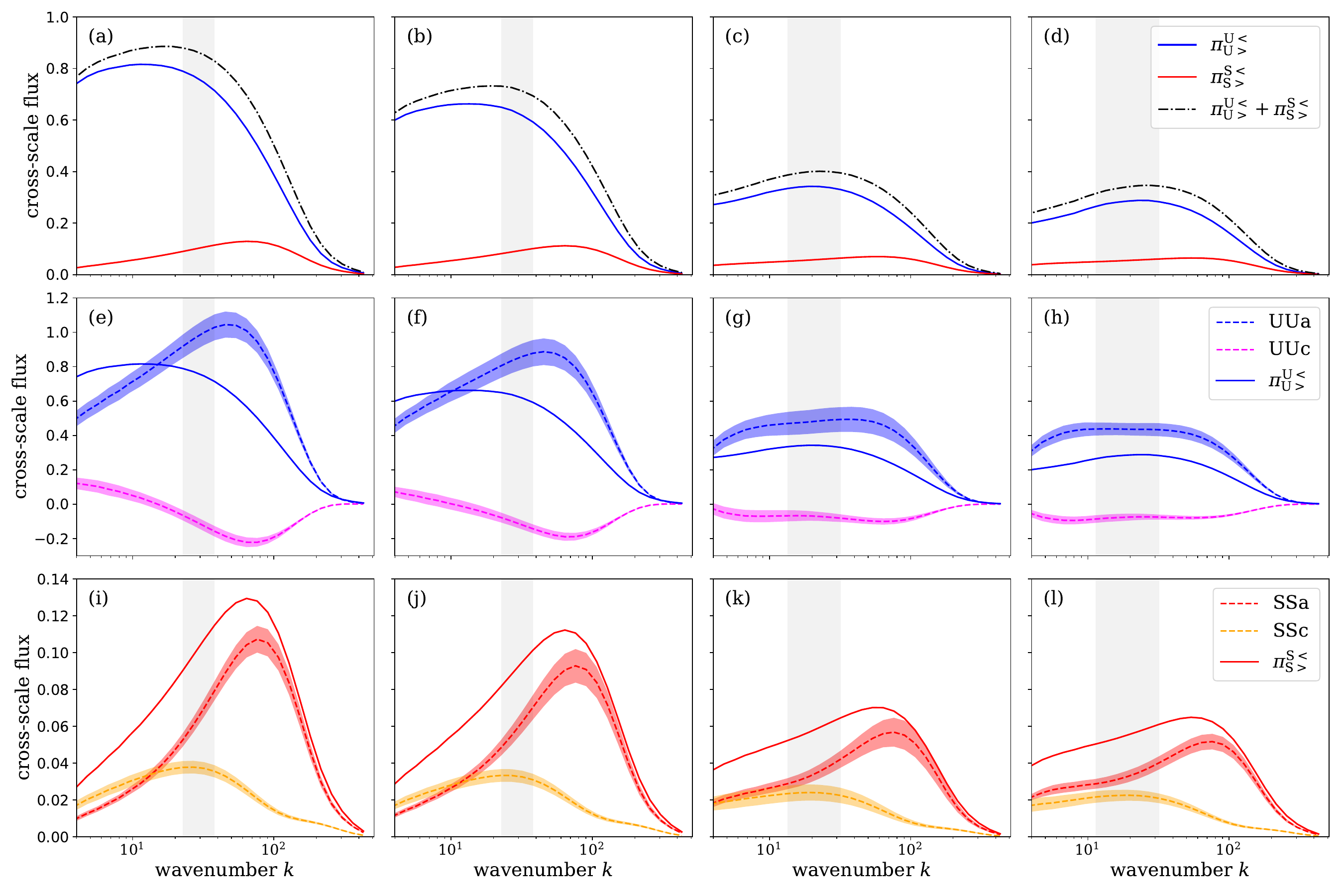}
\caption{Cross-scale fluxes for supersonic turbulence ($V=4.0$) with decreasing fraction of solenoidal forcing modes ($\zeta=2/3$ for (a,e,i), $\zeta=1/2$ for (b,f,j), $\zeta=1/4$ for (c,g,k), and $\zeta=1/8$ for (d,h,l)) as in Fig.~\ref{fig:cross_scale_2to1}.}
\label{fig:cross_scale_vel4}
\end{figure*}

\subsection{Cross-scale flux}
\label{sec:crossFlux}

The total flux of energy across a given wavenumber is defined by equation~\eqref{eq:CrossFluxDef}. In this section, we investigate the dependence of the kinetic and internal energy flux on the characteristic velocity $V$ and the mixture of solenoidal and compressive forcing modes specified by $\zeta$. We do not consider the inter-budget cross-scale fluxes $\Pi^{\mathrm{U}^<}_{\mathrm{S}^>} (k)$ and $\Pi^{\mathrm{S}^<}_{\mathrm{U}^>} (k)$, which produce only minor contributions to the total energy flux (we confirmed this in representative cases; see also the above discussion of the corresponding transfer functions).

First we consider the dependence of the energy flux and its components on $V$ (i.e.\ the forcing magnitude and the resulting Mach number changes). Figure~\ref{fig:cross_scale_2to1} shows our results for the case $\zeta=2/3$. The most striking trend is that the advective component of the kinetic energy flux $\Pi^{\mathrm{U}^<}_{\mathrm{U}^>} (k)$ changes systematically with the Mach number of the flow (blue dashed lines in panels (e,f,g,h)). For $V=1.0$ ($\langle\mathrm{Ma}^2\rangle^{1/2}\approx 1.7$), the advective component is nearly constant. For lower Mach number ($\langle\mathrm{Ma}^2\rangle^{1/2}\approx 0.9$), it decreases slightly with wavenumber, while there is a steep increase for RMS Mach numbers above $3$ (see Table~\ref{tab:simulations}). This is partially compensated by the negative compressive component. As shown in panels (a,b,c,d), the range of wavenumbers with energy transfer around zero is not only narrow, but is displaced from the maximum cross-scale flux in the case of the highest Mach number.\footnote{This is not, as it might seem, a contradiction because the the total energy transfers defined by equation~\eqref{eq:TransfShellK} do not directly correspond to derivatives of the cross-scale fluxes~\eqref{eq:CrossFluxDef}, which are obtained by integrating over a subregion $\shellQ \leq k$ and $\shellK > k$ in the $\shellQ\shellK$-plane.} This suggests a break-down of inertial range scaling toward high Mach numbers for a fixed fraction of solenoidal forcing.

Figure~\ref{fig:cross_scale_vel4} (e,f,g,h) unravels a trend with the forcing parameter $\zeta$: From forcing dominated by solenoidal modes ($\zeta=2/3$) to mostly compressive forcing ($\zeta<1/2$), the advective component of $\Pi^{\mathrm{U}^<}_{\mathrm{U}^>} (k)$ becomes flatter and nearly constant over the inertial-range wave numbers inferred from the transfer functions. In the case $\zeta=1/4$, the gray shaded region in the plot is centered at the peak of the total energy flux (c). In other words, inertial-range properties are restored at higher Mach numbers if the fraction of compressive forcing modes is sufficiently high. There is an influence of the RMS Mach number as well (see also the discussion of density PDFs in Section~\ref{sec:stat} and Table~\ref{tab:simulations}), but the changes are larger than what could be expected solely on the basis of slightly lower Mach numbers for decreasing $\zeta$. Moreover, there appears to be a relatively sharp transition between $\zeta=1/2$ (steep increase of advective component, narrow and displaced inertial subrange) and $\zeta=1/4$ (flat advective component, rather broad inertial subrange around peak of total energy flux). In the next section, we will show that these observations become also manifest in the turbulence energy spectra.

\subsection{Turbulence energy spectra}
\label{sec:numericalSpect}

The energy spectrum function is analytically defined as derivative of the cumulative energy up to a given wavenumber $k$. Formally, this corresponds to shell energies in the limit of infinitesimally thin shells (i.e. surface integrals). Since shell energies can be evaluated from numerical data only for shells of finite thickness $\Delta\shellK$, we define the energy spectrum on the basis of the
shell energy as
\begin{equation}
E_w(k)\Delta\shellK \simeq \Ekin^\shellK
\end{equation}
In addition to $E_w(k)$, we also compute energy spectra for the primitive variable $\V{u}$,
\begin{equation}
E_u(k)\Delta\shellK \simeq
\int_\shellK \frac{1}{2} \FT{u}_i \FT{u}_i^* \dd\V{k}\;,\\
\end{equation}
and for the variable $\V{v}=\rho^{1/3}\V{u}$ introduced by \citet{Kritsuk2007}: 
\begin{equation}
E_v(k)\Delta\shellK \simeq
\int_\shellK \frac{1}{2} \FT{v}_i \FT{v}_i^* \dd\V{k}\;.\\
\end{equation}

\begin{figure*}[!htbp]
\centering
\includegraphics[width=\textwidth]{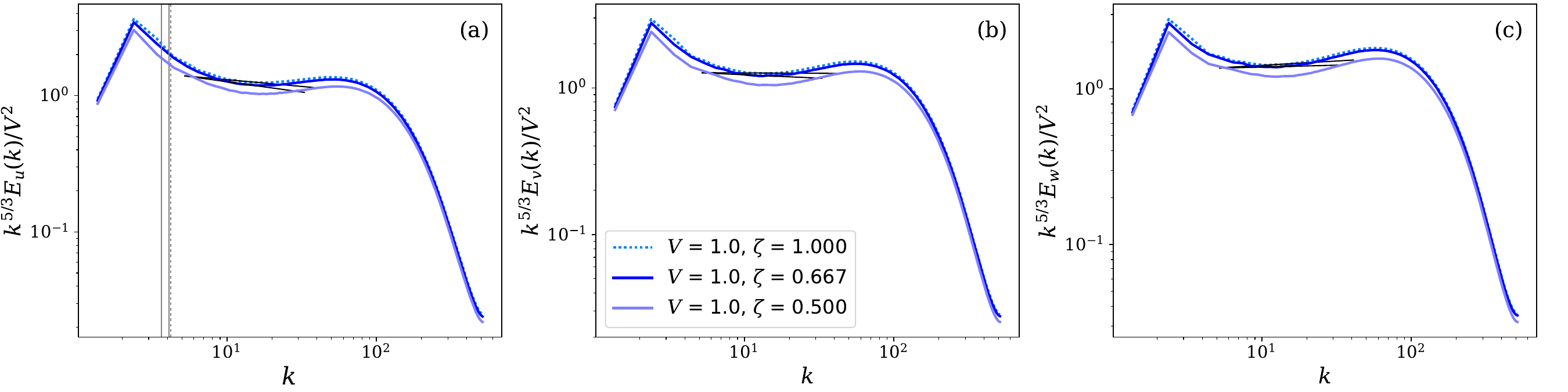}
\includegraphics[width=\textwidth]{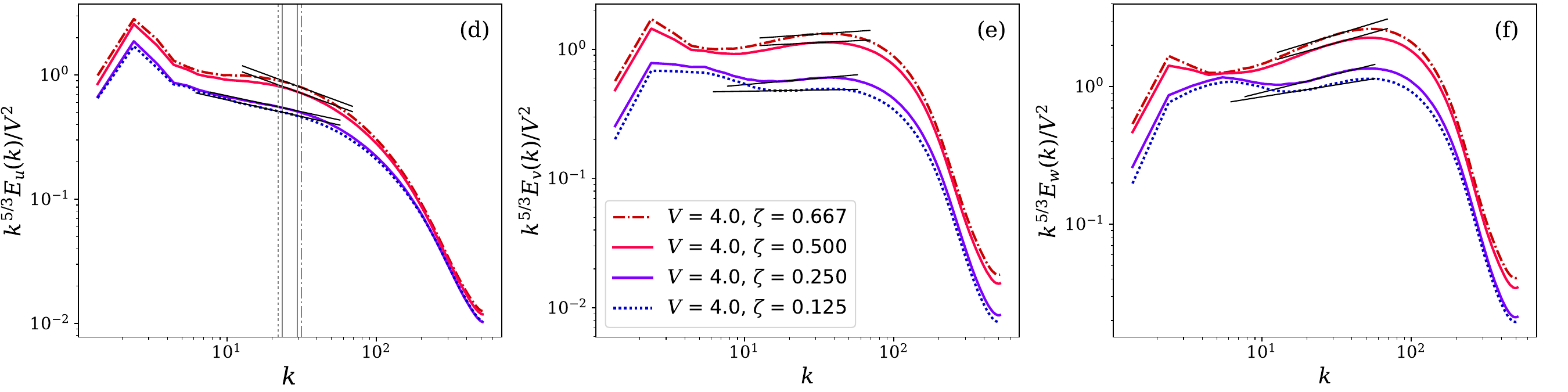}
\caption{Normalized turbulence energy spectra for the three velocity variables $\V{u}$ (a,d), $\V{v}=\rho^{1/3}\V{u}$ (b,e), and $\V{w}=\sqrt{\rho}\,\V{u}$ (c,e) compensated with $k^{5/3}$. The mixture of solenoidal and dilatational forcing modes is varied for integral velocity scales $V=1.0$ (a,b,c) and $V=4.0$ (d,e,f). The power-law fits listed in Table~\ref{tab:powerlaw} are shown as thin black line segments for $0.5k_{\rm min} < k < 2k_{\rm max}$. The vertical gray lines in the left plots indicate the sonic wave numbers $k_{\rm s}$ defined by equation~(\ref{eq:k_sonic}).}
\label{fig:spect_zeta}
\end{figure*}

The three definitions of the turbulence energy spectrum convey different information. The spectrum of pure velocity modes is given by $E_u(k)\propto k^{-5/3}$ in the weakly compressible case, while $E_u(k)\propto k^{-2}$ is the expected scaling in shock-dominated flow at high Mach numbers. This trend is indeed observed for the exponents $\beta_u$ of power-law fits $E_u(k)\propto k^{\beta_u}$ in the range of wavenumbers inferred from the transfer functions (see Table~\ref{tab:powerlaw}). For example, the Kolmogorov slope $\beta_u \approx -1.67$ is recovered in the subsonic case ($V=0.25$) with purely solenoidal forcing and a slope $\beta_u \approx -1.93$ close to Burgers scaling is found for Mach numbers of about $5$ ($V=4.0$) with mainly compressive forcing ($\zeta=1/4$ and $1/8$). In Fig.~\ref{fig:spect_zeta}, sample spectra are plotted for $V=1.0$ (a,b,c) and $V=4.0$ (d,e,f). In the high-Mach case, the strong impact of the forcing parameter $\zeta$ is palpable. As observed for the cross-scale fluxes, the spectrum functions for $\zeta > 1/2$ are markedly different from those with a stronger solenoidal component. In the latter case the estimated inertial range shown as thin black line segments is implausible, as it falls within the bottleneck bump at high wavenumbers. It appears that the bottleneck effect becomes even more pronounced for lower Mach numbers. This might explain why the energy transfer fails to vanish at intermediate wavenumbers in some cases, such as $V=1.0$ and $\zeta=2/3$. In this case, the inferred inertial subrange shrinks to zero and no power law fits are applied (the corresponding simulations are thus omitted in Table~\ref{tab:powerlaw}).

\begin{figure*}[!htbp]
\centering
\includegraphics[width=\textwidth]{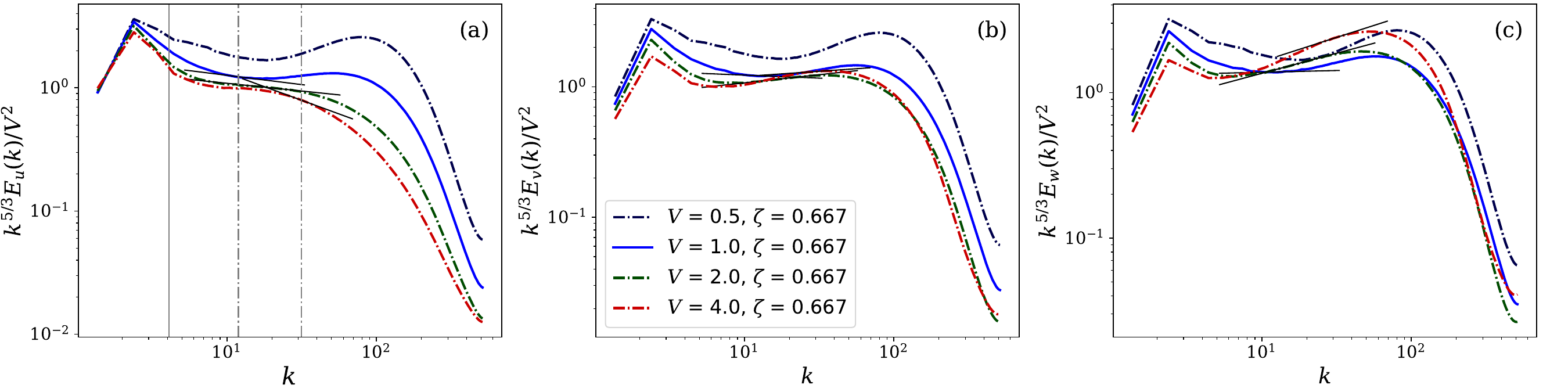}
\caption{Normalized compensated turbulence energy spectra as in Fig.~\ref{fig:spect_zeta} for different Mach numbers and $\zeta=2/3$.}
\label{fig:spect_vel}
\end{figure*}

For each spectrum, we also show the sonic wavenumber $k_{\rm s}$, which is implicitly defined by\footnote{Here we use the initial value of $c_{\rm s}$, which is close to the mean speed of sound for $\alpha=1000$.}
\begin{equation}
    \label{eq:k_sonic}
    \int_{k_{\rm s}}^\infty 2E_u(k)\dd k = c_{\rm s}^2\;,
\end{equation}
i.e.\ the velocity fluctuation exceeds the speed of sound for $k<k_{\rm s}$ \citep{Schmidt2009,Federrath2010}. Although the inertial-range fits for $V=1.0$ cover wavenumbers well above $k_{\rm s}\approx 4.2$, the relatively steep slopes in Table~\ref{tab:powerlaw} reflect a significant impact of compressibility effects. The spectra plotted in Fig.~\ref{fig:spect_vel} show that $k_{\rm s}$ is shifted with increasing $V$ toward higher wavenumbers in the inertial subrange, thus indicating the transition from moderately compressible to supersonic turbulence. Comparing to Fig.~\ref{fig:transf_alpha}, however, one can see that compression-related transfers $\T{SSc}^\shellK$, $\T{SU}^\shellK$, and $\T{US}^\shellK$ have peaks at wave numbers significantly above $k_{\rm s}$. As a consequence, wavenumbers for which compression effects are relatively strong are not limited by the sonic wavenumber.

Motivated by the dimensional expression $\rho u^3/L=v^3/L$ for energy flux, it was suggested that the incompressible $k^{-5/3}$ scaling can be extended into the compressible regime for the variable $\V{v}$ \cite{Kritsuk2007}. The fits to $E_v(k)$ listed in Table~\ref{tab:powerlaw} show that $\beta_v$ is close to $-5/3$ for solenoidal forcing at low to moderate Mach numbers ($V\le 1.0$). At higher Mach numbers, it appears that $\beta_v=-5/3$ can be reached for increasingly compressive forcing, although the results for $V=4.0$ are not fully conclusive. 
The corresponding spectra are shown in Figs~\ref{fig:spect_zeta} (b,e) and~\ref{fig:spect_vel} (b). Similar to the cross-scale energy flux discussed in the previous section, we see a transition of the shape of $E_v(k)$ for $V=4.0$ when the forcing becomes dominated by compressive modes. 
A prominent feature for $\zeta=1/4$ and $1/8$ is the shift of the kink at which the forcing peak joins into the flat part of the spectrum to wavenumbers above 10. It turns out that this is also the lower bound of the fit range (solely established on the basis of the transfer functions). For higher $\zeta$ (mainly solenodial forcing), these two points move apart and the spectra become flatter (see Fig.~\ref{fig:spect_zeta} and corresponding fit parameters in Table~\ref{tab:powerlaw})

\begin{table}[tbp]
        \begin{tabular}{llrrccc}
                \toprule
                $V$ & $\zeta$ & 
                $k_{\rm min}$ & $k_{\rm max}$ &
                $\beta_{u}$ & $\beta_{v}$ & $\beta_{w}$  \\
\colrule
0.25 & 1 & 9.5 & 45.2 & -1.667 & -1.663 & -1.661 \\ %
\colrule
1.0 & 1 & 9.5 & 22.6 & -1.766 & -1.672 & -1.611 \\ %
1.0 & 2/3 & 9.5 & 19.0 & -1.826 & -1.721 & -1.651 \\ %
\colrule
2.0 & 2/3 & 9.5 & 32.0 & -1.792 & -1.542 & -1.391 \\ %
2.0 & 1/2 & 9.5 & 32.0 & -1.772 & -1.528 & -1.387\\ %
2.0 & 1/4 & 9.5 & 22.6 & -1.867 & -1.684 & -1.584\\ %
\colrule
4.0 & 2/3 & 22.6 & 38.1 & -2.110 & -1.587 & -1.337 \\ %
4.0 & 1/2 & 22.6 & 38.1 & -2.101 & -1.609 & -1.363  \\ %
4.0 & 1/4 & 13.5 & 32.0 & -1.925 & -1.578 & -1.402 \\ %
4.0 & 1/8 & 11.3 & 32.0 & -1.936 & -1.647 & -1.492 \\ %
\botrule
        \end{tabular}
        \caption{Power-law fits for statistically stationary turbulence in the estimated inertial range $[k_{\rm min}, k_{\rm max}]$ following from the criteria formulated in Section~\ref{sec:numericalTransfer} (gray shaded wavenumber ranges in Fig.s~\ref{fig:transf_alpha}, \ref{fig:cross_scale_2to1}, and~\ref{fig:cross_scale_vel4}). The slopes of the energy spectra $E_u(k)$, $E_v(k)$, and $E_w(k)$ are given by the exponents $\beta_u$, $\beta_v$, and $\beta_w$, respectively. For details about the fitting procedure see Appendix~\ref{sec:compute}.}
        \label{tab:powerlaw}
\end{table}

The spectrum function $E_w(k)$ corresponds to the kinetic energy density of compressible density (see Section~\ref{sec:ShellEn}). It has previously been noted that $E_w(k)$ tends to be flatter than both $E_u(k)$ and $E_v(k)$ \cite{Kritsuk2007,Federrath2010,Grete2017a}, possibly approaching $k^{-4/3}$ in the strongly supersonic case. Qualitatively, the plots in Fig.~\ref{fig:spect_zeta} (c,f) indicate that $E_w(k)$ becomes flatter than the reference $k^{-5/3}$ spectrum for high Mach number ($V=2.0$ and $V=4.0$). As can be seen from the power-law fits listed in Table~\ref{tab:powerlaw}, $\beta_w$ is closer to $-4/3$ if the forcing is dominated by solenoidal modes. This trend can also be seen in the plot showing $E_w(k)$ for fixed $\zeta=2/3$ in Fig.~\ref{fig:spect_zeta}  (c). In view of our analysis of energy transfer and cross-scale fluxes, it follows that a slope of $-4/3$ for the energy spectrum function $E_w(k)$ is incommensurate with an inertial-range cascade. 

\section{Conclusions}
\label{sec:concl}

We investigated energy transfer functions and the resulting cross-scale fluxes of kinetic and internal energy in implicit large eddy simulations of forced compressible turbulence. To maintain a statistically stationary state even at high Mach numbers, the net increase of internal energy caused by the dissipation of kinetic energy is compensated by a simple linear cooling function. We varied both the amplitude of the forcing, resulting in Mach numbers ranging from subsonic to supersonic, and the mixture of solenoidal and compressive modes. Indicators of nearly isothermal turbulence are approximately constant RMS Mach numbers and log-normal density PDFs with significant skewness  in the highly compressible regime \cite{Federrath2010,Federrath2013,Hopkins2013a}.  

To compute transfer functions we apply a spectral decomposition into shells based on the variables $w=\sqrt{\rho}\,u$ (kinetic energy transfer) and $q=\sqrt{\rho}\,\cs$ (internal energy transfer), where $u$ and $\cs$ are the gas velocity and speed of sound, respectively. For a numerical resolution of $1024^3$, we find only limited regimes of vanishing energy transfer, which is a defining property of an inertial subrange. In a resolution study (see Appendix~\ref{sec:resolution}) we find that extended regimes of vanishing transfers and, thus, an inertial subrange just begins to emerge for $1024^3$ grid cells. For the majority of simulations, the total transfer function crosses zero at intermediate wavenumbers and we consider shells to be close to inertial-range scaling if the modulus of the total transfer is less than $10\,\%$ of the peak value at higher wavenumbers.

For the comparison of the kinetic and internal energy budgets, cumulative quantities are better suited than the transfer functions. Cross-scale fluxes quantify the total amount of energy that is passed from all shells below a given wavenumber into shells above that wavenumber (we do not consider the question of interaction locality of energy transfer here). Our analysis shows that the cross-scale flux is always dominated by kinetic energy. This is of course expected in the weakly compressible case, where density fluctuations contribute only little to the energy transfer between different shells. Since the gas is nearly isothermal, changes in $q$ do not play a significant role. The flux of internal energy becomes more important at transonic Mach numbers, but decreases relative to the kinetic energy flux toward higher Mach numbers. This 
is simply a consequence of the large fluctuation of the velocity compared to the speed of sound. 
Moreover, the small transfers between the kinetic and internal energy reservoirs, which are related to pressure dilatation, support the theoretical reasoning in \cite{Aluie2011}. As a consequence, it appears that the kinetic energy cascade  becomes asymptotically invariant in the limit of high Mach numbers.

We also find a strong sensitivity of the energy transfers on the composition of the 
forcing for supersonic turbulence similar to the previously reported sensitivity of density PDFs or energy spectra \citep{Federrath2010}.
Particularly the advective component of the kinetic energy flux turns out to be nearly constant for intermediate wavenumbers if the compressive fraction of the force field is sufficiently large. At lower Mach numbers, roughly constant kinetic energy flux is only found for mostly solenoidal forcing. Together with the compressive component of the kinetic energy flux and the internal energy flux, the behaviour is less clear. A simple statistical argument suggests a ratio of two to one for the power of solenoidal and compressive large-scale modes (corresponding to $\zeta=1/2$; see \citep{Federrath2010}) in the strongly compressible regime. Although our results indicate that there is a preferred mixture of solenoidal and compressive forcing, it appears that the ratio changes depending on the Mach number. This is probably a consequence of the increasingly strong coupling of solenoidal and compressive modes. To disentangle the coupling and its effect on the inertial range, further separation of the transfers into divergence-free and rotation-free modes is required.

The impact of the forcing mixture is corroborated by power-law fits to turbulent energy spectra in a range that is solely based on the transfer functions. Although the resolution is barely sufficient to resolve inertial-range scaling (for high Mach numbers, the fit range is shifted toward the bottleneck range in which the spectrum functions are tilted), we can nevertheless discern some trends. The slopes $\beta_u$ of the pure velocity spectra $E_u(k)$ show the expected dependence on the Mach number. While $\beta_u\approx -5/3$ for the lowest Mach number, the slope approaches $-2$ (Burgers turbulence) for higher Mach numbers (if the forcing is mainly solenoidal, the slope even falls below $-2$). We do not find a clear separation of scalings below and above the sonic wavenumber, as proposed in \cite{Galtier2011}. Based on our analysis, the inertial range crosses the sonic wavenumber and the slope gradually changes with increasing Mach number. However, as mentioned above, this also entails a stronger bias by the bottleneck effect. The compressible energy density spectrum $E_w(k)$ tends to become shallower with increasing Mach number, with $\beta_w$ between about $-4/3$ and $-5/3$ depending on the forcing. The dependence of the slope $\beta_v$ for the variable $v=\rho^{1/3}u$ on the weighing of solenoidal and compressive modes deserves closer attention. In \cite{Kritsuk2007}, the Kolmogorov value of $-5/3$ was proposed for $\beta_v$ as a universal scaling exponent. With increasing Mach number, we find that $\beta_v$ tends to be closer to $-5/3$ for increasingly compressive forcing. In contrast to \cite{Federrath2010,Federrath2013}, we do not find a much steeper slope provided that the range of power-law fits is constrained by vanishing transfer functions (e.g.\ $\beta_v=-2.1$ for $\langle\mathrm{Ma}^2\rangle^{1/2}\approx 5.5$ and $\zeta=0$ in \cite{Federrath2010} was obtained for a wavenumber range that is more strongly affected by the forcing term $ \mathcal{F}^\shellK$ in the shell-energy equation~(\ref{eq:RateEkin_shell}) than in the range we obtain from our transfer analysis). Taken by themselves, the spectrum functions $E_v(k)$ do not exclude steeper fits. Galtier \& Banerjee \cite{Galtier2011} indeed favor a broader definition of compressible turbulence, with a scale-dependent compressible contribution to the energy flux in addition to a scale-invariant major component that can be identified with the advective component of the kinetic energy flux. Interpreted in this theoretical framework, our findings indicate that inertial-range scaling in the more specific sense of nearly constant energy flux is limited to a certain region in parameter space in which the net compressible contribution is relatively small. A direct comparison to observational data is difficult because the full three-dimensional structure of the velocity and density fields has to be reconstructed. For example, observations of star-forming clouds favor scaling exponents in between the Kolmogorov and Burgers exponents \cite{Federrath2010,Roman2011}. By using indicators such as density PDFs, some evidence has been found that the driving mechanism resembles mixed forcing \cite{Kainul2013}.

It has to be stressed that turbulence in a periodic box driven by stochastic forcing is an idealistic model. For turbulent flows occurring in nature, the actual mode of energy injection and realistic boundary conditions and maybe even initial conditions have to be taken into account. In the case of astrophysical turbulence, gravity, magnetic fields, and highly non-linear cooling increase the complexity even further.
Moreover, microphysical effects such as deviations from local thermodynamic equilibrium are expected to play a role. However, this is beyond the scope of our simple model (as implied by the EOS
and cooling function) and is left for more advanced studies in the future.
Our numerical study demonstrates that even under the most ideal circumstances (statistically homogeneous and isotropic flow in a steady state), compressible turbulence exhibits a rich phenomenology. Apart from the incompressible limit ($q\gg w$), the notion of a compressible inertial-range cascade probably has to be regarded as asymptotic property in the limit of high Mach numbers ($q\ll w$). However, compressible turbulence in astrophysical systems is often found in the intermediate regime ($q\sim w$). An important example is the intracluster medium in clusters of galaxies \cite{Miniati2015a,Schmidt2016}.

\bigskip 

\section*{Acknowledgments}

We thank Ann Almgren (LBNL) for supporting us in the implementation of stochastic forcing into the Nyx code.
PG acknowledges funding by NASA Astrophysics Theory Program grant \#NNX15AP39G.
The simulations presented in this article were performed on the Cray XC (TDS) system of the North-German Supercomputing Alliance (HLRN) facilities in Berlin. 
We also acknowledge the yt toolkit by \citet{Turk2011} that was used for our analysis of numerical data.

%

\appendix

\begin{figure*}[!htbp]
\centering
\includegraphics[width=0.48\textwidth]{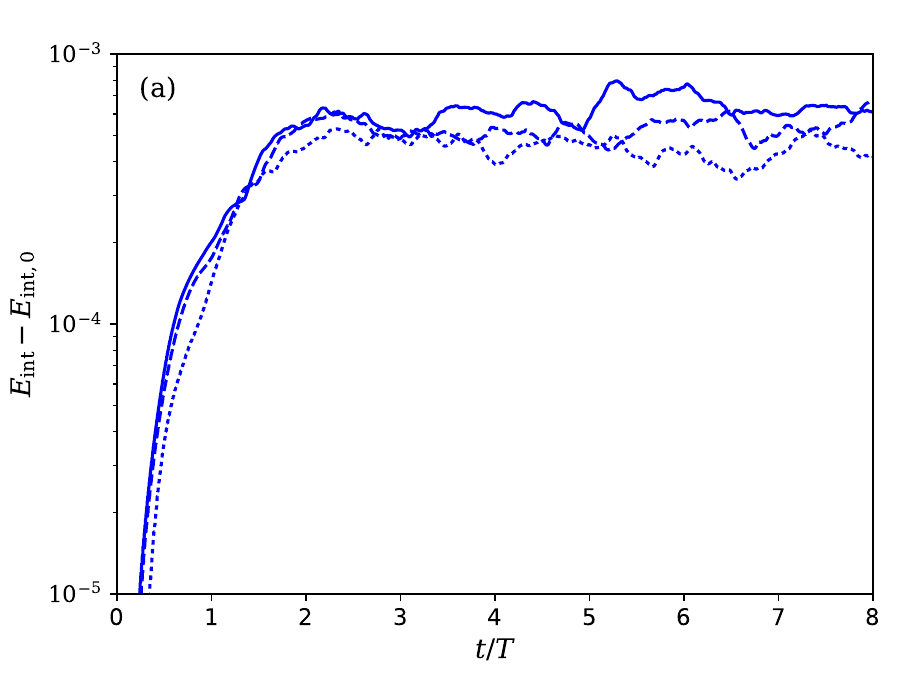}
\includegraphics[width=0.48\textwidth]{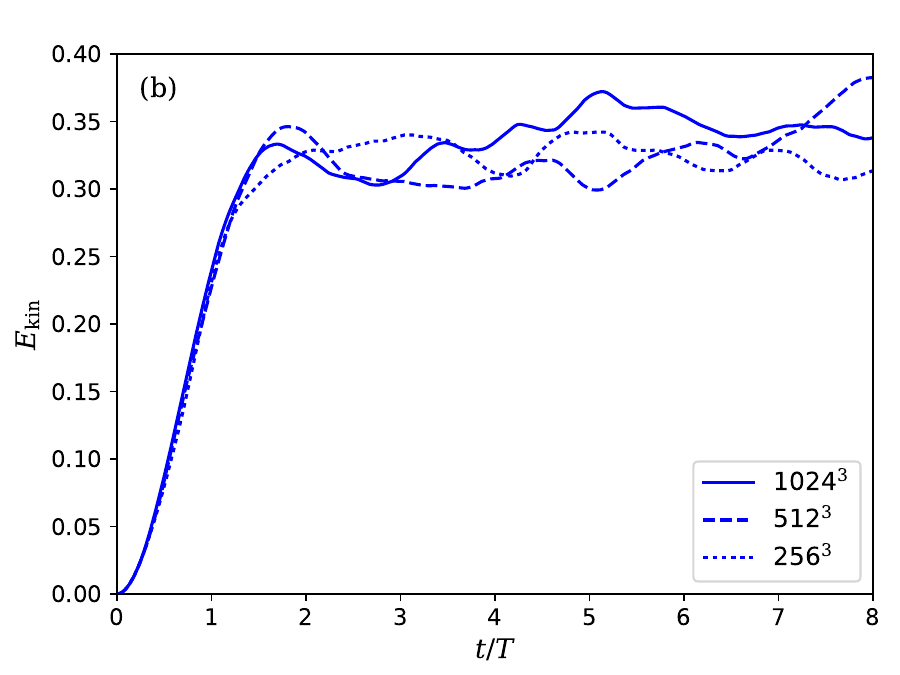}
\caption{Time evolution of the mean internal energy minus initial energy (a) and kinetic energy (b) for different numerical resolutions in the case $V=1.0$ and $\zeta=2/3$.}
\label{fig:mean_res}
\end{figure*}

\section{Computational aspects}
\label{sec:compute}

Nyx uses a hierarchical parallelisation scheme based on the AMReX framework\footnote{See \href{https://amrex-codes.github.io/amrex/}{https://amrex-codes.github.io/amrex}}. The grid is decomposed into a suitable number of boxes, which are distributed among MPI tasks. Each box in turn is split into several tile boxes \cite{Zhang2016}, which extend over the full size of a box in $x$-direction (corresponding to the innermost loops in the Fortran core routines of the code) and covering quadratic tiles of size $16\times 16$ in the $yz$-plane. Each OMP thread computes one or more tile boxes. This scheme is also beneficial for efficient vectorization. For high performance, it is of particular importance to enable vectorization for the inverse Fourier transform of the forcing spectrum to spatial grid, which is basically a multiply-add operation over an array with more than 100 elements for each grid cell (corresponding to the number of non-zero Fourier modes; the spectrum of the forcing is exactly as specified in \cite{Schmidt2009}). Source terms (forcing and heating/cooling) are treated with a Strang splitting method which is described in detail in \cite{Almgren2013}. The full code including the implementation of stochastic forcing is publicly available on Github.\footnote{AMReX repo: \href{https://github.com/AMReX-Codes/amrex}{github.com/AMReX-Codes/amrex},\\
Nyx repo: \href{https://github.com/AMReX-Astro/Nyx}{github.com/AMReX-Astro/Nyx}.}

The simulations presented in this article were performed on Xeon Phi (Knights Landing) processors with AVX-512 vectorization. We used 32 nodes for the largest grids with $1024^3$ cells. On each node, 8 MPI tasks ran with 32 OMP threads per task on 64 physical cores (i.e.\ we employed the maximal number of hyperthreads, resulting in a speedup by a factor of about $2$ compared to no hyperthreading). The nodes were operated in cache mode with Sub-NUMA-2 clustering. With this configuration, the wallclock time per CFL timestep was about $7$ seconds. The total number of timesteps required for a sufficient number of dynamical time scales varied between roughly 20000 and 40000.

The postprocessing was carried out with Python, using parallel Fourier transforms implemented in the \texttt{mpi4py-fft} package\footnote{See \href{https://mpi4py-fft.readthedocs.io/en/latest/}{mpi4py-fft.readthedocs.io}.} to compute transfer functions. To read data cubes from the simulations, we used the \mbox{AMReX} frontend implemented in \texttt{yt}\footnote{See \href{http://yt-project.org/doc/}{yt-project.org/doc}.} \cite{Turk2011}.

To fit power laws to energy spectra, we applied the \texttt{scipy} library function \texttt{curve\_fit} to logarithmic data, i.e.\ the data model is $y = \beta x + y_0$ with $x = \log k$ and $y = \log E(k)$. Since our basis hypothesis is that the energy transfer vanishes in the inertial range, we assume the uncertainty \texttt{sigma} of the data in \texttt{curve\_fit} to be proportional to the deviation of the energy transfer from zero. As a result, shells with energy transfer close to the upper bound of $0.1$ relative to the peak value have less weight in relation to the shell in which the transfer crosses zero.

\begin{figure*}[!htbp]
\centering
\includegraphics[width=\textwidth]{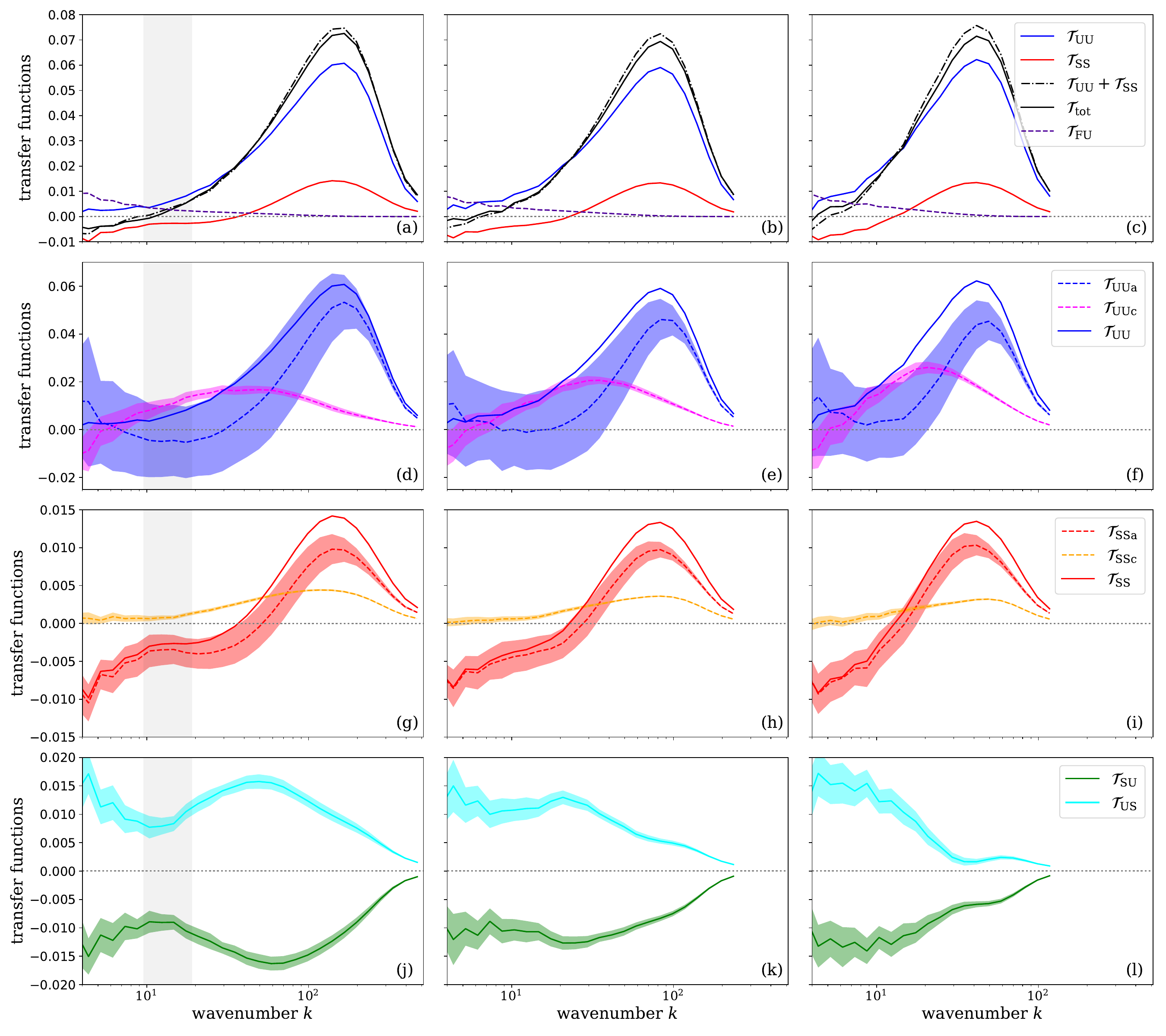}
\caption{Time-averaged transfers as in Fig.~\ref{fig:transf_alpha} for numerical resolutions $1024^3$ (a,d,g,j), $512^3$ (b,e,h,k), and $256^3$ (c,f,i,l) in the case $V=1.0$ and $\zeta=2/3$.
In panels (a,b,c) also the transfer term associated with the forcing, $\mathcal{F}^\shellK \equiv \T{FU}^\shellK$, is shown for comparison.}
\label{fig:transf_res}
\end{figure*}

\begin{figure*}[!htbp]
\centering
\includegraphics[width=\textwidth]{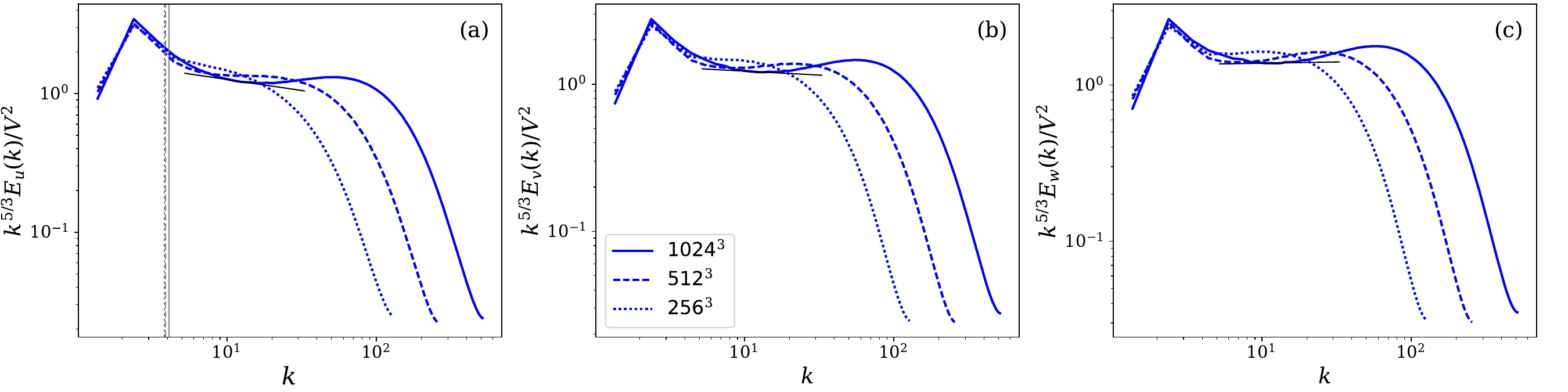}
\caption{Normalized compensated turbulence energy spectra as in Fig.~\ref{fig:spect_vel} for different numerical resolutions in the case $V=1.0$ and $\zeta=2/3$.}
\label{fig:spect_res}
\end{figure*}

\section{Resolution study}
\label{sec:resolution}

Figure~\ref{fig:mean_res} shows the time evolution of the mean internal (a) and kinetic  (a) energies for $V=1.0$ and $\zeta=2/3$. For this particular case we performed simulations with $256^3$, $512^3$, and $1024^3$ grid cells. Within statistical variations due to the randomness of the forcing, the energies saturate at comparable levels. This implies that dissipation and cooling are insensitive to numerical resolution (see also \cite{Schmidt2006,Schmidt2011}).

The resolution dependence of the total transfer functions is shown in Fig.~\ref{fig:transf_res}. In this case, we also computed the energy transfer due to the forcing, $\mathcal{F}^\shellK \equiv \T{FU}^\shellK$ in eq.~\eqref{eq:RateEkin_shell} (see also \cite{Grete2017a}). Figure~\ref{fig:transf_res} (a,b,c) shows that the contribution of forcing is significant in comparison to non-linear kinetic energy transfer for wavenumbers lower than about $20$. For small wave numbers, the forcing dominates and $\T{tot}^\shellK\simeq -\T{FU}^\shellK$ (this can be seen in panel (a) for the highest resolution). The kinetic and internal energy transfer functions, $\T{UU}^\shellK$ (d,e,f) and $\T{SS}^\shellK$ (g,h,i), are qualitatively similar. However, an important feature that clearly changes with resolution is the location of the positive peaks. Particularly for the advective components, the peaks are shifted to the left as the range of numerically resolved wavenumbers decreases. This demonstrates that these peaks are coupled to the numerical dissipation scale, as explained in Section~\ref{sec:numericalTransfer}. The compressive components, on the other hand, are less affected by the grid resolution (at least for resolutions $512^3$ and $1024^3$), which suggests a physical origin of the maximum. For resolutions below $1024^3$, the total transfer is positive down to low wavenumbers. As a consequence, there is no wavenumber interval qualifying as inertial subrange according to the criterion defined in Section~\ref{sec:numericalTransfer}. This is also reflected by the transfers $\T{SU}^\shellK$ and $\T{SU}^\shellK$ (j,k,l), which are relatively weak in the inertial subrange for the highest resolution. For $512^3$, the minima (of the absolute values) at $k\approx 10$ are only marginal and they disappear completely for the lowest resolution. This supports the proposition of \cite{Aluie2011} that pressure-dilation effects causing exchange of kinetic and internal energy should be weak in the inertial subrange.

The resulting energy spectrum functions are plotted in Fig.~\ref{fig:spect_res}. The dissipative cutoff of the spectra shows the expected dependence on the numerical resolution. Even at intermediate wavenumbers, the spectra hardly converge as an inertial subrange only begins to emerge for $1024^3$ grids. This highlights the difficulty of applying power-law fits based on the appearance of the spectrum functions. One could certainly determine fits also for the lower resolutions. Depending on the chosen wavenumber range, this would result in significantly different slopes. It can also be seen that the bottleneck effect tends to become stronger with increasing resolution. These difficulties in establishing inertial range scalings were also reported in \cite{Kritsuk2007,Federrath2013} for resolutions up to $4096^3$.

\end{document}